\documentclass{article}
\usepackage{authblk}
\usepackage{graphicx}
\usepackage[table]{xcolor}
\usepackage{multirow}
\usepackage{amsmath}
\usepackage{cite}

\begin{document}
\title{New static structures in the strained carbon and boron chains}
\date{}
\author{G.M. Chechin\thanks{gchechin@gmail.com}~}
\author{V.S. Lapina}
\affil{Southern Federal University, Research Institute of Physics, Zorge~5, Rostov-on-Don, Russia}

\maketitle

\begin{abstract}

We discuss bi-structures with one or two long bonds in the central part of the strained monoatomic chains appearing abruptly as a result of hard bifurcations of the static form with increasing strain above some critical value. Structures of this type were initially revealed in the model of the monoatomic chains with the Lennard-Jones interactions (L-J model). There are well-defined arguments in favor of the fact that the above bi-structures are universal in the sense that they can exist in the finite strained chains of any physical nature. We tested this hypothesis on the strained chains of carbon and boron atoms with the aid of the density functional theory (DFT), using results obtained in the simulation of the L-J model as initial approximation. The properties of these bi-structures, depending on the length of the chains, are investigated in detail. It is assumed that the abrupt change in the electrical properties of strained carbon and boron chains in the vicinity of the bifurcations can be used to create nanodevices within the framework of straintronics.
\end{abstract}

\section{Introduction} \label{Seq1}

In recent years, there has been a steadily growing interest in studying of carbynes representing one-dimensional carbon structures~\cite{1}.

A comprehensive review of various physical, chemical and mechanical properties of carbynes, methods of their synthesis and research with the aid of spectroscopic methods can be found in Ref.~\cite{2}. Below, we discuss only those properties of carbyne, which are directly related to our present research and the works that are not reflected in the above review.

\subsection{Monoatomic carbon chains and their properties \label{Seq1.1}}

Monoatomic carbon chains (carbynes) can exist in two different forms. These are cumulene with double bonds between all atoms [chemical structure ($=C=C=)_{n}$] and polyyne with the alternation of single and triple bonds [chemical structure $(-C\equiv
C-)_{n}$]. As a result, all bond lengths (BLs) are identical in cumulene, while polyyne demonstrates alternation of short and long bonds. 

Carbynes possess many unique properties. In particular, they are the strongest material known at the present time (carbyne has a tensile strength twice that of graphene and Young's modulus three times greater than graphene), cumulene is a conductor better than linear gold chain and can be considered as the thinnest nanowire. An energy gap arises in the electron spectrum of the cumulene with an appropriate strain, and it changes from the conductor to a semiconductor or dielectric, etc.

It is assumed that the unique carbyne properties will allow using this material in future for various purposes of nanotechnology. In particular, carbyne chains are interesting because they can connect different fragments of graphene and can be used as the smallest nanowire in metal-matrix nanocomposites. Carbyne is considered as a promising material for spintronics, straintronics, as well as for hydrogen storage in hydrogen technology, etc. 

It is interesting that there are IR spectroscopy evidences of the existence of carbyne in the inter-stellar dust clouds. However, the synthesis of carbyne in the laboratory and the experimental study of its properties present great difficulties~\cite{2}. So far it was possible to synthesize free carbon chains consisting of only a few dozen atoms. It was reported that the chain of 6000 carbon atoms was obtained, but inside a carbon nanotube~\cite{3}.

Free carbon chains of any length must be terminated by molecular complexes to ensure their stability. In principle, they can be obtained with two different terminations: $sp^{3}$ termination, resulting in carbon atoms linked by alternated single and triple bonds (polyyne) and, thus, with alternating bond lengths, and $sp^{2}$ termination resulting in double bonds (cumulene). Note that experimentally polyyne is found to be more stable than cumulene.

Particular interest represents the study of {\it oriented} carbyne, which is an ensemble of short carbon chains perpendicular to the substrate surface to which they are attached with their hydrogen ends~\cite{4,5,6,7}. In particular, it is argued in~\cite{7} that this material can be a topological insulator with two-dimensional superconductivity. 

\subsection{Static properties of carbon chains in the framework of the density functional theory \label{Seq1.2}}

As was already noted, the chemical synthesis of carbynes and their experimental study encounters great technical difficulties and, therefore, theoretical methods are very important. Most of these methods are based on the density functional theory (DFT)~\cite{8,9,10}, implemented in a number of powerful computational packages, such as Abinit, Quantum Espresso, VASP, and others. Many interesting results were obtained in this way for infinite and finite carbon chains~\cite{11,12,13,14,15,16,17,18,19,20,21,22}.

In Ref.~\cite{15}, for long strained carbyne chains with even number of atoms, the Peierls phase transition was predicted above a certain threshold of the strain. As a result of this transition, carbyne transforms from metallic state to insulator state, and one can use this property in nanodevices to control the conductivity of the material by mechanical action. 

Properties of sufficiently short carbon chains are discussed in~\cite{11,12,13,17,18}.

In Ref.~\cite{12} DFT methods allowed investigating in detail the so called ``parity effect'' for the strained carbon chains of finite size. It means that the distribution of bond lengths and magnetic moments at atomic sites exhibit even-odd disparity depending on the number of carbon atoms in the chain. The authors of this paper also studied the dependence of BLs on the type of saturation of carbon chains at their both ends. Several hydrogen atoms can be attached to their ends to passivate chemically active ends of the chains. If two hydrogen atoms are attached to each end of the chain, then the bond lengths in its middle part correspond approximately to the cumulene structure. If only one hydrogen atom is attached to each end of the chain, then the polyyne structure appears. This problem, as well as the methods for identifying various forms of carbyne by infrared absorption spectra are discussed in detail in Ref.~\cite{2}.

In Ref.~\cite{12}, it was also found that local perturbation created by a small displacement of the single carbon atom at the center of a long chain induces oscillations of atomic forces and charge density, which are carried to long distances over the chain. 

\subsection{Nonlinear atomic vibrations in strained carbon chains and our approach for finding new static structures in these objects \label{Seq1.3}}

In Ref.~\cite{21}, large amplitude atomic oscillations in the strained carbon chains were studied with the aid of DFT modeling and a sharp  {\it softening} of the $\pi$-mode frequency was found above a certain critical strain value. Condensation of this mode also leads to the Peierls transition discussed in~\cite{15}. Moreover, the soft mode concept allowed the authors to suggest, that there can exist two new forms of carbyne, which differ from the polyyne in the  {\it type of alternation} of short and long chemical bonds. 

In the same paper, a simple classical model was proposed which allows explaining the above softening of nonlinear normal modes at certain values of the strain. This model represents a monoatomic chain whose interparticle interactions are described by the Lennard-Jones potential. Hereafter we refer to it as the L-J model. Thus, in Ref.~\cite{12} we begin with DFT modeling and then use L-J chain to interpret the obtained results with the aid of this simple mechanical model.

In the present paper, we move in the opposite direction. Indeed, we begin with investigation of static structures in the strained L-J chains and only then use the obtained results in order to choose an adequate initial configuration for its refinement with the aid of DFT modeling. Such approach is required due to the fact that the potential energy of the strained carbon chain within the framework of the DFT model is a multi-extremal function and the choice of the initial approximation determines to which  {\it local minimum} of this function we get as a result of using the DFT descent method. 

This approach has allowed us to reveal new static structures that were not found in the works of our predecessors~\cite{23,24}.

\subsection{The structure of the present paper \label{Seq1.4}}

In Sec.~\ref{Seq2}, we consider the L-J model used to study the properties of strained monoatomic chains, as well as some computation details. Appearance of the bi-structure with one long bond in the strained L-J chains, as a result of a rigid bifurcation, is discussed in Sec.~\ref{Seq3}. Similar bi-structures in the strained carbon and boron chains obtained by DFT-modeling are considered in Secs.~\ref{Seq4} and ~\ref{Seq5}. Section~\ref{Seq6} is devoted to investigation of the bi-structure with two long bonds. Here we also study strained carbon and boron chains with the aid of DFT-modeling. In Sec.~\ref{Seq7} we present some additional discussions and summarize results of the present paper.

\section{Mathematical models of strained monoatomic chains and some computational details \label{Seq2}}

In the present paper, we use two different models to study the static properties of monatomic chains. The first one is the ``Lennard-Jones model'', which represents an equidistant monoatomic chain whose interparticle interactions are described by the Lennard-Jones potential: 

\begin{equation}
\label{eq1}
\varphi(r)=\frac{A}{r^{12}}-\frac{B}{r^{6}}. 
\end{equation}

Here, $r$ is the distance between two particles, while $A$ and $B$ are phenomenological parameters. These parameters can be set equal to unity  {\it without loss of generality} ($A$ = 1, $B$ = 1). It is possible to do this by the appropriate scaling of the time and spatial coordinates in Newton's equations, which describe dynamics of the considered system. We refer to the Lennard-Jones potential (\ref{eq1}) with $A$ = 1, $B$ = 1 as the standard L-J potential. The force of interaction corresponding to the standard L-J potential has the form:

\begin{equation}
\label{eq2}
f(r)=-\frac{d\varphi }{dr}=\frac{12}{r^{13}}-\frac{6}{r^{7}}.
\end{equation}

We call the Lennard-Jones model applied to the chain of $N$ identical particles as the ``L-J chain''. As was already noted in Introduction, this simple model occurs to be very successful in explaining dynamics of carbon chains calculated in the framework of the density functional theory.

The Lennard-Jones model belongs to the class of models considered in the molecular dynamics approach, in which the particles are considered as mass points whose dynamics is described by the classical Newton equations.

In the present work, all calculations of L-J chains were performed by the homemade programs written in the Maple package~\cite{25}. 

The second model used in our paper is the "DFT-model" corresponding to the density functional theory. We analyze the properties of strained chains of carbon and boron atoms with the aid of this model. Using the term "DFT-model", we mean that practical calculations within the framework of the density functional theory require a number of different  {\it approximations}. These approximations are implemented in the software packages ABINIT, Quantum Espresso, VASP, etc. 

For our purpose, Quantum Espresso and ABINIT were used with the following traditional approximations:
\begin{itemize}
\item The Born-Oppenheimer approximation, which allows one to separate the slow motion of ions relative to the fast motion of electrons;

\item Local density approximation (LDA) for the exchange-correlation functional;

\item Pseudopotentials by Troullier and Martins (MT) or by Perdew, Burke, Ernzerhof (PBE)~\cite{26}.

\item The basis of plane waves for solving the Kohn-Sham equation with the cutoff equal to 450 eV;

\item Energy self-consistency accuracy was $10^{-8}$ eV;

\item 	Force accuracy for finding ionic configurations was $10^{-4}$ eV / $\buildrel_\circ \over {\mathrm{A}}$.
\end{itemize}

The DFT model is a significantly more adequate approximation to the real atomic systems, since it takes into account the presence of the electron shells of atoms, which are polarized during ions' movement, and this effect is taken into account with the aid of the quantum-mechanical approach. 

The self-consistent solution of the Kohn-Sham equations determine the state of the electronic subsystem, which adjusts to the ion configuration. Then the Gel'man-Feynman forces acting on the ions are found, and a certain time step is taken to solve the classical Newton's equations for ionic subsystem. The above process is repeated until reaching the self-consistency, in which all the forces are zero with a given degree of accuracy. Actually, this procedure represents a descent method (usually the conjugate gradient method) to minimize the potential energy of the system.

 Since the energy is, in general case, a multi-extreme function in the configuration space, the choice of the initial approximation (the initial configuration to be refined by DFT minimization) is of fundamental importance. Indeed, the energy minimum to which the descent method will lead us depends on the choice of the initial approximation. An essential point of our work is that we find the initial approximation for DFT-modeling taken into account the results of the study of static structures of the Lennard-Jones model. Namely, such approach allows us revealing several new static structures of monatomic chains of carbon and boron atoms.

\section{Appearance of the bi-structure with one long bond in strained Lennard-Jones chains \label{Seq3}}

Here we consider the bi-structure with one long bond (bi-structure 1) in the strained monoatomic chains with interparticle interaction described by the Lennard-Jones potential (L-J chains). As was just mentioned, we use the results of studying L-J chains to choose an appropriate initial approximation for refining the static structures of the strained carbon chains in the framework of the density functional theory.

\begin{figure}[h!] \centering
	\includegraphics[width=1\linewidth]{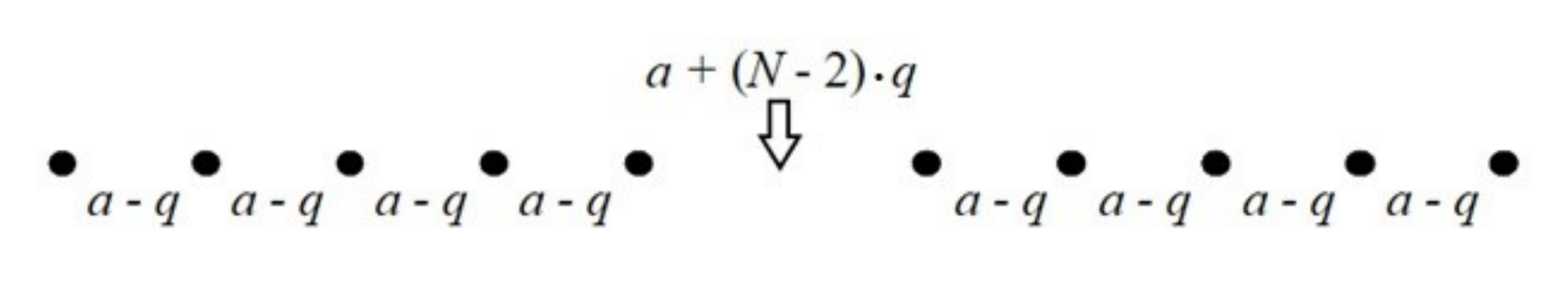} \caption{The bi-structure with one long bond in the strained $N$-particle L-J chain.} \label{fig1}
\end{figure}

We look for the static structure of the strained N-particle L-J chain in the form shown in Fig.~\ref{fig1}.

This structure contains $N-2$ short bonds $(a-q)$ and one long bond $[a+(N-2)\cdot q]$, where a is the interparticle distance in the equidistant strained L-J chain, while $q > 0$ determines the reduction of the parameter $a$ when bi-structure 1 appears. Actually, the structure in Fig.~\ref{fig1} represents two identical equidistant {\it subchains} with interparticle distance $(a-q)$, which are spaced apart by the distance equal to the long bond $[a+(N-2)\cdot q]$. 

Note that if we consider atomic displacements from the equidistant structure of the strained L-J chain before bifurcation, then the displacements corresponding to the bi-structure form the {\it exact arithmetic progression} with a difference $q$ in each subchains.

Since we look for the state of the chain in {\it equilibrium}, the Lennard-Jones forces, acting on each atom from the left and right neighbors, should be equal. For the atoms located at the junction of {\it two short bonds}, this condition is satisfied automatically due to the equality of their lengths. Therefore, it is enough to require the equilibrium condition of atoms located only at the junction of the short and long bonds, and therefore the following equation holds, which hereafter we refer to as ``force equation'':

\begin{equation}
\label{eq3}
f(a-q)=f[a+(N-2)\cdot q]
\end{equation}

Here $a$ is the interatomic distance of the strained equidistant L-J chain, while arguments of the forces on the l.h.s. and on the r.h.s. of this equation are the short bond $(a-q)$ and the long bond, respectively. We must solve Eq.~\ref{eq3} for an unknown $q$, which shows by how much the short length is shorter than $a$. 

\begin{figure}[h!] \centering
	\includegraphics[width=1\linewidth]{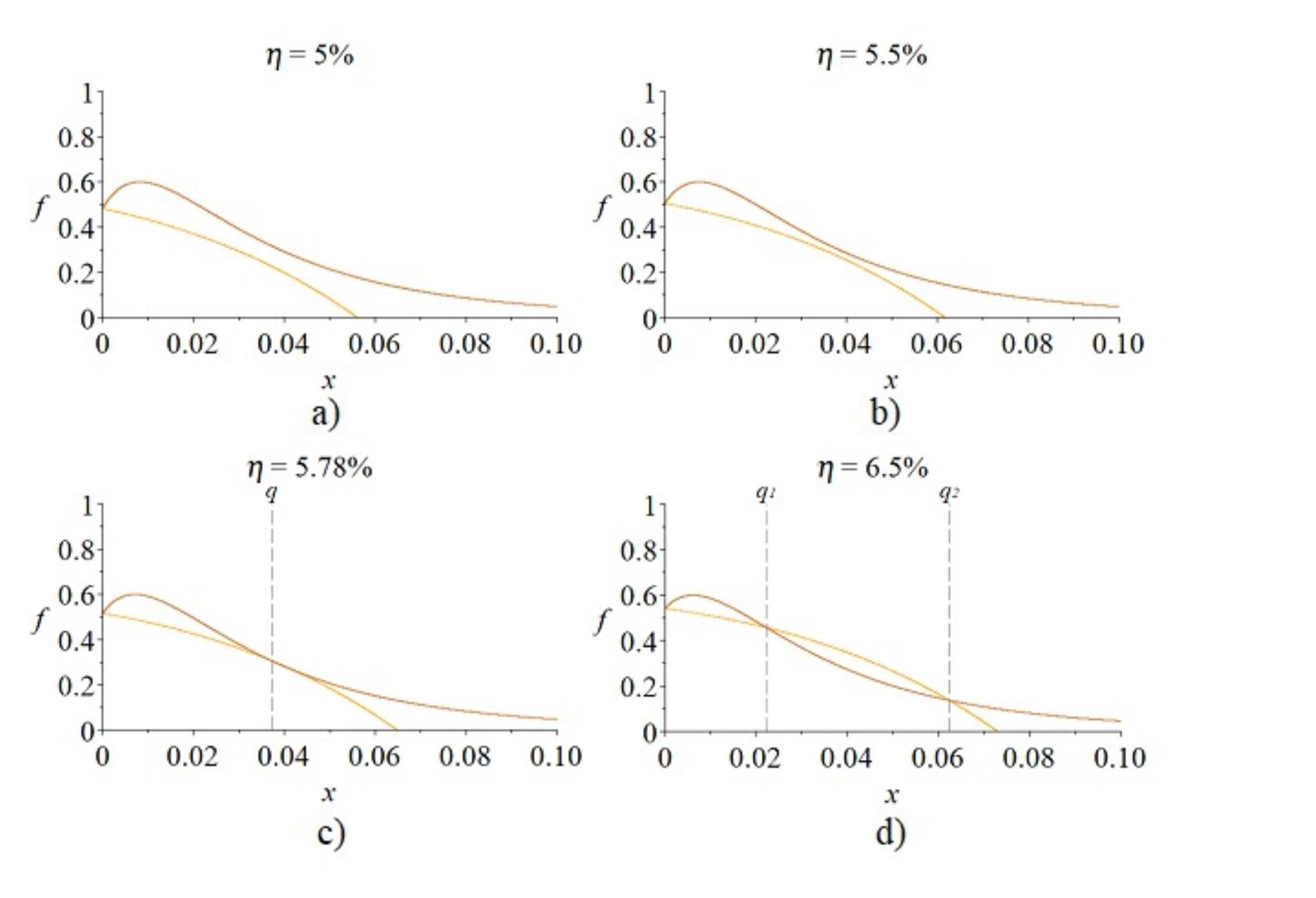} \caption{Graphical solution of the force Eq.~\ref{eq3} for the L-J chain with N=8 particles.} \label{fig2}
\end{figure}

Equation~\ref{eq3} can be represented as a non-linear algebraic equation of high degree, which has several real and complex roots. These roots can be easily calculated with the aid of Maple. Among these roots, we are interested only in the value of $q$, which has a physical meaning and leads to the appearance of the stable bi-structure with short bonds $(a-q)$ and one long bond $[a+(N-2)\cdot q]$. This root arises as a result of a $rigid$ bifurcation that occurs with increasing strain $\eta$, and it is convenient to illustrate the process of its occurrence by means of the graphical method. For this purpose, we depict in Fig.~\ref{fig2} the left and right side of the force equation~\ref{eq3} for different values of the L-J chain strain. It can be seen that for $\eta = 5\,{\%}$ and $5.5\,{\%}$ the above plots do not overlap, but tend to approach each other. In Fig.~\ref{fig2}c, these plots touch each other $(\eta=5.78\,{\%})$, and the point of tangency corresponds to the appearance of a new real root $q$ as a result of a rigid bifurcation, while two new roots, $q_{1}$ and $q_{2}$, appear with further increase of $\eta$.

The root $q_{1}$ corresponds to the maximum, while $q_{2}$ corresponds to the minimum of the total potential energy of the L-J chain. Therefore, $q_{2}$ determines the lengths of short and long bonds in the stable bi-structure in the strained L-J chain because the static state must correspond to the minimum of the system potential energy.

In Fig.~\ref{fig2}, the energy profiles $U(q)$ of the L-J chain for different values of the strain are presented. From this figure, one can see the appearance of the above new roots $q_{1}$ and $q_{2}$ of the force equation~\ref{eq3}, which correspond to maxima and minima of the potential energy, as well as their evolution with increase of the strain of the chain with $N=8$ particles.

\begin{figure}[h!] \centering
	\includegraphics[width=1\linewidth]{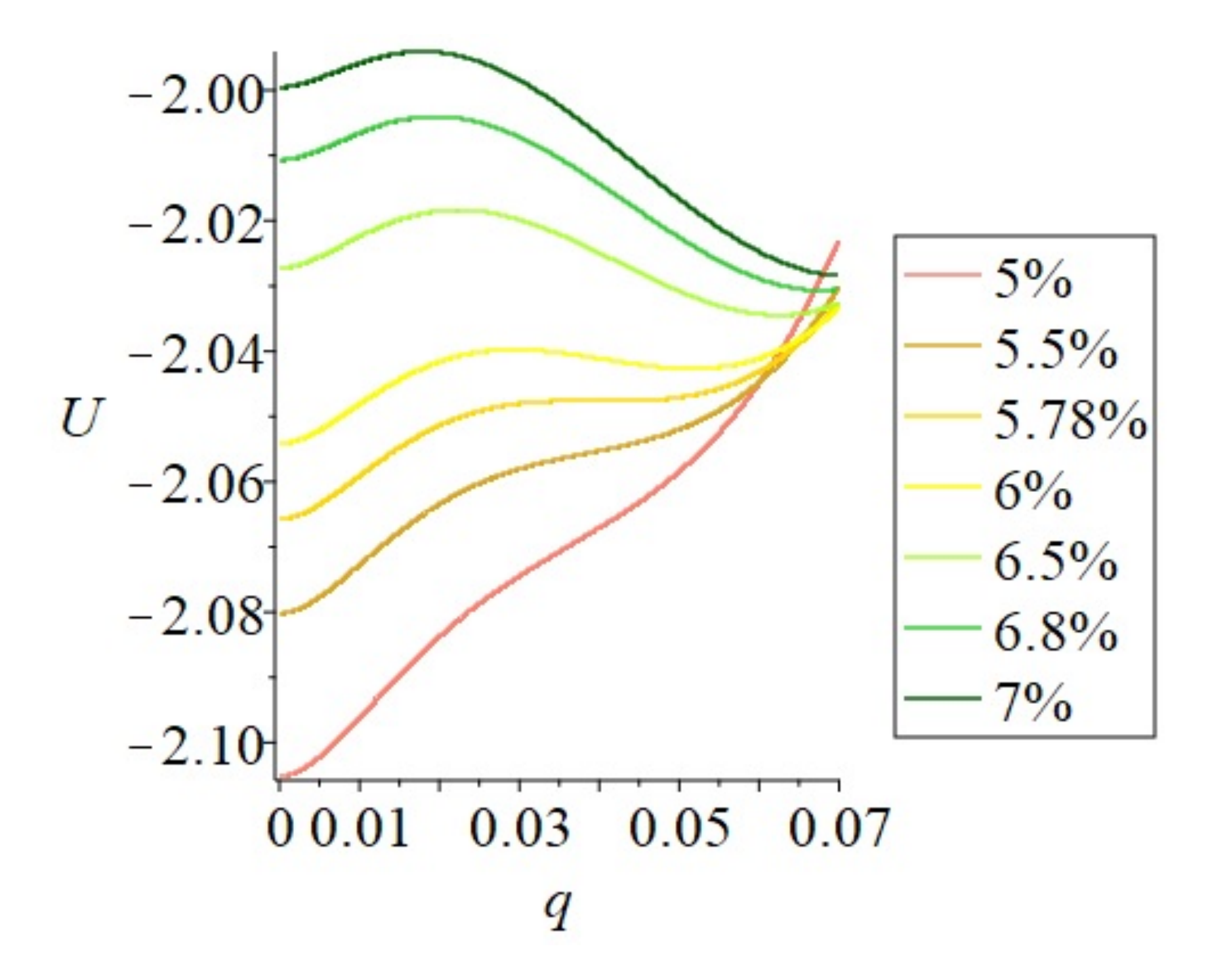} \caption{Evolution of the energy profiles $U(q)$ for the L-J chain with $N=8$ particles for different strains (color online).} \label{fig3}
\end{figure}

\begin{table}
	\caption{Dependence of the root $q=q_{2}$ of the force equation (Eq.~\ref{eq3}) on the number $N$ particles in the chain for the strain $\eta=10\,{\%}$.}
	\renewcommand{\tabcolsep}{0.018\linewidth}
	\label{t1}
	
	\begin{tabular}{ccccccccc}
		\hline
		$N$ & 8 & 12 & 16 & 20 & 40 & 60 & 80 & 100 \\
		\hline
		$q_{2}$ & 0.1071 & 0.1112 & 0.1119 & 0.1121 & 0.1122 & 0.1123 & 0.1123 & 0.1123 \\
		\hline
	\end{tabular}
\end{table}

It can be seen from this table that the root $q=q_{2}$ of Eq.~\ref{eq3} slightly depends on the number of particles $N$ in the L-J chain. The critical strain $\eta_{c}$ and the bond lengths of the bi-structure appearing after the above rigid bifurcation depend also on $N$. The parameters of the bi-structure with one long bond discussed in this section are presented in Table~\ref{t2}.

\begin{table}\centering
	\caption{Dependence of the critical strain $\eta_{c}$ on the number of atoms $N$ in the L-J chain.}
	\renewcommand{\tabcolsep}{0.02\linewidth}
	\label{t2}
	{\rowcolors{2}{lightgray}{white} 
	\begin{tabular}{cccc}
		\hline
		$N$ & Critical strain $\eta_{c},\,{\%}$ & Long bond & Short bond \\
		\hline
		6   &	8.05 &	1.284 &	1.165 \\
		\hline
        8   &	6.70 &	1.308 &	1.154 \\
        \hline
        10  &	5.78 &	1.334 &	1.146 \\
        \hline
        12  &	5.10 &	1.351 &	1.142 \\
        \hline
        14  &	4.58 &	1.365 &	1.139 \\
        \hline
        16  &	4.17 &	1.379 &	1.137 \\
        \hline
        18  &	3.84 &	1.394 &	1.135 \\
        \hline
        20  &   3.56 &  1.404 & 1.134 \\
        \hline
        22  &	3.32 &	1.408 &	1.134 \\
        \hline
        24  &	3.12 &	1.421 &	1.132 \\
        \hline
        26  &	2.95 &	1.436 &	1.131 \\
        \hline
        28  &	2.79 &	1.438 &	1.131 \\
        \hline
        34  &	2.42 &	1.458 &	1.130 \\
        \hline
        44  &	2.00 &	1.486 &	1.128 \\
        \hline
        54  &	1.72 &	1.519 &	1.127 \\
        \hline
        104 &	1.05 &	1.610 &	1.125 \\
		\hline
	\end{tabular}}
\end{table}

The argument of the force on the r.h.s. of Eq.~\ref{eq3} is a long bond, which for large $N$ corresponds to the {\it tail part} of the Lennard-Jones potential and, therefore, this force can be quite small. For the case $N = 20$ and $\eta=10\,{\%}$, it is equal to $f=0.0016$. Therefore, the left side of Eq.~\ref{eq3} also must be small. However, the Lennard-Jones force on the finite interval of its argument can be small only near the minimum of the potential (it is zero at this minimum itself). Thus, in the case of a large long bond, the force equation~\ref{eq3} is associated with two different regions of the Lennard-Jones potential, namely, with the tail and the vicinity of its minimum.

The above arguments allow us to estimate bond lengths in the bi-structure for large values of $N$. Indeed, both left and right sides of Eq.~\ref{eq3} tend to zero when $N\rightarrow  \infty$. Therefore, for this limiting case $f(a-q)=0$, where $a=a_{0}\cdot (1+\eta)$ and $a_{0}=1.12246$. It follows from this condition that $a-q=a_{0}$, since the Lennard-Jones force is zero at the point $a_{0}$ corresponding to the potential minimum. Then we have $a_{0}\cdot (1+\eta)-q=a_{0}$ and, as a result, $q=a_{0}\cdot \eta$. On the other hand, the data in Table~\ref{t1} are given for the strain $\eta=10\,{\%}$ and, therefore, $q=1.12246\cdot0.1=0.112246$. This value of the root $q$ coincides with its limiting value given in Table~\ref{t1}.

In Table~\ref{t3} we present short and long bonds for $N=20$ and different values of the strain $\eta$.

\begin{table}\centering
	\caption{Parameters of the bi-structure with one long bond for L-J chains with $N= 20$ particles for different values of the strain $\eta$.}
	\renewcommand{\tabcolsep}{0.02\linewidth}
	\label{t3}
	
	\begin{tabular}{ccc}
		\hline
		Strain  $\eta,\,{\%}$ & Long bond & Short bond \\
		\hline
		4  &	1.8866 	&	1.1274 \\
\hline
5  &	2.1537	&	1.1244 \\
\hline
6  &	2.3849	&	1.1234 \\
\hline
7  &	2.6061	&	1.1230 \\
\hline
8  &	2.8233	&	1.1228 \\
\hline
9  &	3.0387	&	1.1226 \\
\hline
10 &	3.2532	&	1.1226 \\
\hline
11 &	3.4672	&	1.1225 \\
\hline
12 &	3.6809	&	1.1225 \\
\hline
13 &	3.8944	&	1.1225 \\
\hline
14 &	4.1078	&	1.1225 \\
\hline
15 &	4.3212	&	1.1225 \\

		\hline
	\end{tabular}
\end{table}

Above we have mentioned that our method of the bi-structure construction is based on the analysis of the properties of the force equation~\ref{eq3}, which connects the Lennard-Jones forces in the vicinity of the tail of the potential and in the vicinity of its minimum. Since any physically adequate interatomic potentials have a minimum and a tail, it is clear that the above arguments for appearance of bi-structures are valid for a wide class of potentials. In other words, the bi-structures similar to those in L-J chains can exist in monatomic chains with arbitrary interatomic potentials.

\section{Bi-structure with one long bond in the strained carbon chains obtained by DFT modeling \label{Seq4}}

As was already mentioned, to reveal the state of the carbon chain similar to bi-structures, which we found in the L-J chains, it is necessary to find an adequate initial approximation (initial atomic configuration) to optimize the structure of the chain with the aid of DFT-modeling. The results obtained while studying the L-J chains allow us determining the following initial state of the carbon chain, which leads to the appearance of the bi-structure with one long bond within the framework of the density functional theory.

We take two identical cumulene chains with interatomic distances corresponding to their strained state, $a=a_{0}\cdot(1+\eta)$, and place them on the same straight line at a distance $R$ from each other, which is certainly larger than $a$. For our purpose, it is sufficient to choose $R=3\,\buildrel_\circ \over {\mathrm{A}}$. After DFT-optimization, this distance transforms into the long bond between carbon atoms in the above discussed bi-structure, and this bond occurs {\it shorter} than $R$, i.e. two our chains attract each other.

The main qualitative difference between the bi-structures obtained in the DFT model and those in the L-J chains is that the former are {\it heterogeneous}, in contrast to the homogeneity of the latter. Indeed, in the bi-structure of the L-J chain, all short bonds are strictly identical, and this bi-structure is homogeneous in this sense, while in the case of real monoatomic chains, such as chains of carbon or boron atoms, short bonds are of {\it different} lengths (the longest are bonds at the ends of the chain).

Similar to the L-J chains, the bi-structure 1 in the DFT model, {\it appears} abruptly with increase of the strain, i.e. as a result of a {\it rigid bifurcation}. As an example, let us consider the appearance of the bi-structure 1 in the carbon chain with $N=16$ atoms. Table~\ref{t3} shows the lengths of interatomic bonds directly {\it before} and {\it after} the bifurcation, i.e. after passing through the critical value of the strain $\eta_{c}= 6.10\,{\%}$ (we obtain this value with $0.01\,{\%}$ accuracy). The data presented in this table clearly indicate the rigid type of bifurcation, leading to the appearance of the bi-structure with one long bond in the center of the carbon chain.

The first column contains the numbers of interatomic bonds for $i=1..15$, where $i$ is the bond number between atoms with numbers $i$ and $i+1$.

\newcommand{\specialcell}[2][c]{%
  \begin{tabular}[#1]{@{}c@{}}#2\end{tabular}}
\begin{table}\centering
	\caption{Appearance of the bi-structure with one long bond in the carbon chain with $N=16$ atoms. Bond lengths ($\buildrel_\circ \over {\mathrm{A}}$) before and after bifurcation obtained with the aid of DFT modeling.}
	\renewcommand{\tabcolsep}{0.02\linewidth}
	\label{t4}

	\begin{tabular}{ccc}
		\hline
		 Bond number & \specialcell{Bond length for \\ $\eta=6.10\,{\%}$ \\ (before bifurcation)} & \specialcell{Bond length for \\ $\eta=6.11\,{\%}$ \\ (after bifurcation)} \\
		\hline
1	 & 1.421	& 1.311 \\
\hline                   
2	 & 1.403	& 1.309 \\
\hline                   
3	 & 1.361	& 1.279 \\
\hline                   
4	 & 1.381	& 1.305 \\
\hline                   
5	 & 1.363	& 1.272 \\
\hline                   
6	 & 1.380	& 1.321 \\
\hline                   
7	 & 1.364	& 1.276 \\
\hline                   
\rowcolor{lightgray} 8	 & 1.379	& 2.582 \\
\hline                   
9	 & 1.364	& 1.276 \\
\hline                   
10	 & 1.380	& 1.321 \\
\hline                   
11	 & 1.363	& 1.272 \\
\hline                   
12	 & 1.381	& 1.305 \\
\hline                   
13	 & 1.361	& 1.279 \\
\hline                   
14	 & 1.403	& 1.309 \\
\hline                   
15	 & 1.421	& 1.311 \\
\hline
\hline                   
$E/N, eV$ & -7.60	& -7.40 \\
		\hline
	\end{tabular}
\end{table}

In Fig.~\ref{fig3}, we depict the bond lengths (BLs), which are given in Table~\ref{t4}. The bonds before and after bifurcation are shown by squares and circles, respectively. The scale along the vertical axes is different in Fig.~\ref{fig3}a and Fig.~\ref{fig3}b, while in Fig.~\ref{fig3}c the same bonds are depicted on one scale.

\begin{figure}[h!] \centering
	\includegraphics[width=1\linewidth]{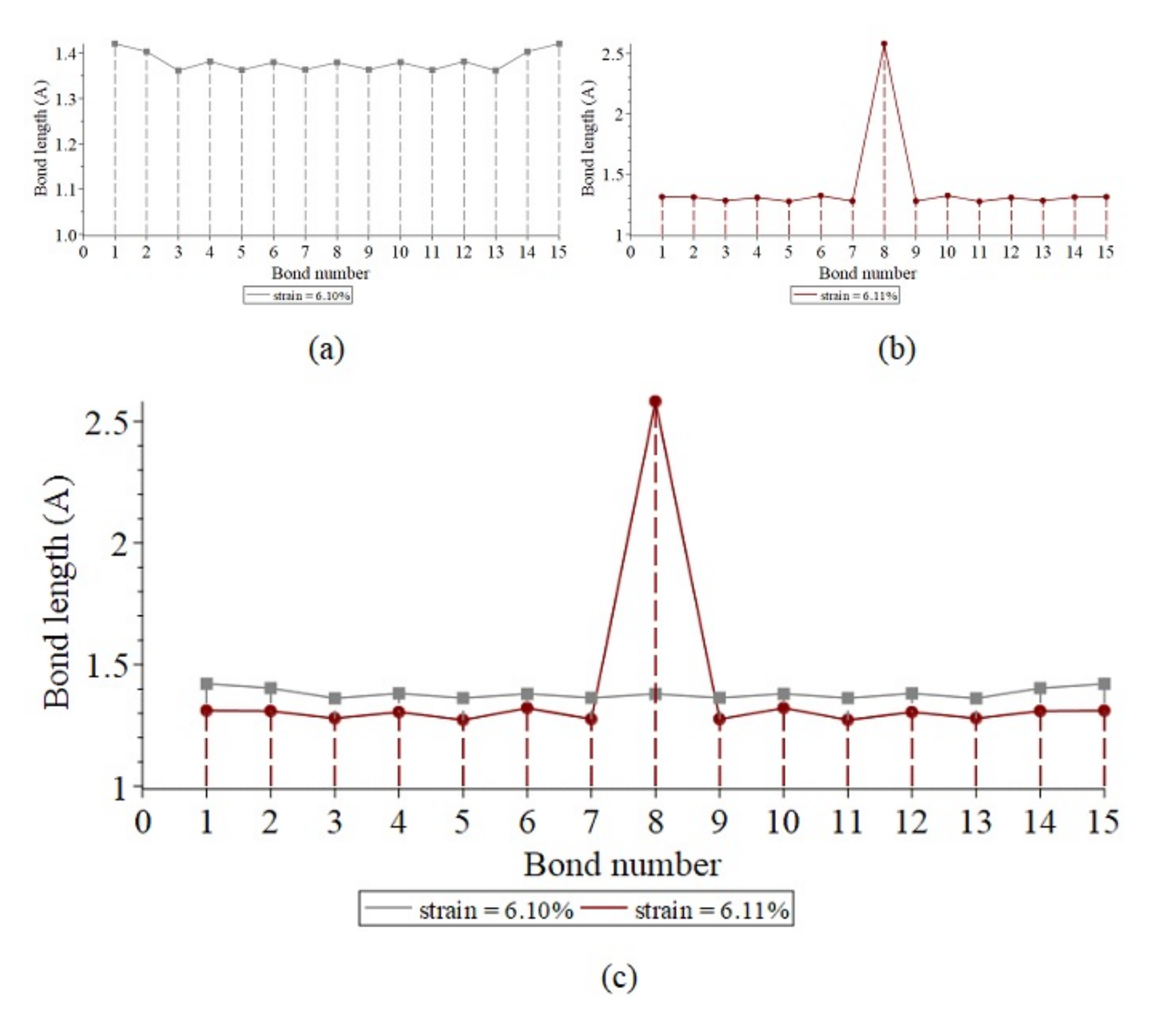} \caption{Bond lengths before (a) and after (b) bifurcation for the chain with $N=16$ atoms. (c) the data of (a) and (b) combined in the same graph (color online).} \label{fig4}
\end{figure}

The above-mentioned heterogeneity of the bi-structures in the carbon chain is clearly visible in this figure. The longest are those interatomic bonds that are located near the ends of the chain, while in the middle there is a slight alternation of bond lengths similar to that of polyyne. 

In the bi-structure 1 resulting from the bifurcation, the long bond separating two equal subchains is approximately twice as large as the values of short bonds (this length depends on the number $N$ of atoms in the chain). Moreover, it can be noted from Fig.~\ref{fig4}c that all short bonds after the bifurcation are reduced compared to those before this bifurcation, e.g. atoms from the center of the chain shift to its ends. 

Note that the static structure {\it before} the bifurcation is similar to that obtained in Refs.~\cite{12,18} and this fact may be considered as an evidence of the correctness of our results. However, the bi-structures arising {\it after} the bifurcation and their further evolution with increase of the strain of the chain are essentially new.

\section{Some properties of bi-structures with one long interatomic bond in the strained carbon chains \label{Seq5}}
\subsection{Dependence of the critical strain of the chain on the number of its atoms and the new parity law \label{Seq5.1}}

The critical strain, above which the bi-structures can exist, and parameters of these bi-structures depend significantly on the number $N$ of atoms forming the chain. We present the corresponding information in Table~\ref{t5}.

\begin{table}\centering
	\caption{Dependence of the critical strain and parameters of bi-structures with one long bond on the number $N$ of atoms in the carbon chains calculated by DFT modeling.}
	\renewcommand{\tabcolsep}{0.02\linewidth}
	\label{t5}

{\rowcolors{4}{lightgray}{white} 
	\begin{tabular}{ccccc}
		\hline
		 \multirow{2}*{$N$} & \multirow{2}*{Critical strain $\eta_{c}, \,{\%}$} & \multirow{2}*{Long bond, $\buildrel_\circ \over {\mathrm{A}}$} & \multicolumn{2}{c}{Short bonds, $\buildrel_\circ \over {\mathrm{A}}$} \\
		 \cline{4-5}
		 & & & Min & Max \\
		\hline
		
8 	& 14.68		& 2.443		& 1.297		& 1.364 \\
\hline
10	& 7.06		& 2.752		& 1.272		& 1.287 \\
\hline
12	& 8.80		& 2.517		& 1.281		& 1.329 \\
\hline
14	& 4.77		& 2.419		& 1.262		& 1.290 \\
\hline
16	& 6.11		& 2.582		& 1.272		& 1.321 \\
\hline
18	& 3.88		& 2.644		& 1.261		& 1.281 \\
\hline
20	& 4.56		& 2.666		& 1.268		& 1.313 \\
\hline
22	& 2.65		& 2.532		& 1.260		& 1.297 \\
\hline
24	& 3.57		& 2.745		& 1.266		& 1.298 \\
\hline
26	& 2.07		& 2.499		& 1.263		& 1.307 \\	
		
		\hline
	\end{tabular} }
\end{table}

It can be seen from this table that there is a clear tendency of decreasing the critical strain $\eta_{c}$ with increase of the number $N$ of carbon atoms. However, the function $\eta_{c}(N)$ is not monotonic and seems rather strange. The reason for this phenomenon is a certain {\it parity effect}, different from that described in~\cite{12,18}. Indeed, in these papers authors discuss the parity effect\footnote{According to the parity effect, many properties of the carbon chains depend significantly on the number N of their atoms, for example, chains with odd N are more durable.}, corresponding to the entire chain of $N$ atoms, while our parity effect is associated with {\it subchains}, each of which is formed by $N/2$ atoms. Let us consider this question in more detail.

\begin{figure}[h!] \centering
	\includegraphics[width=1\linewidth]{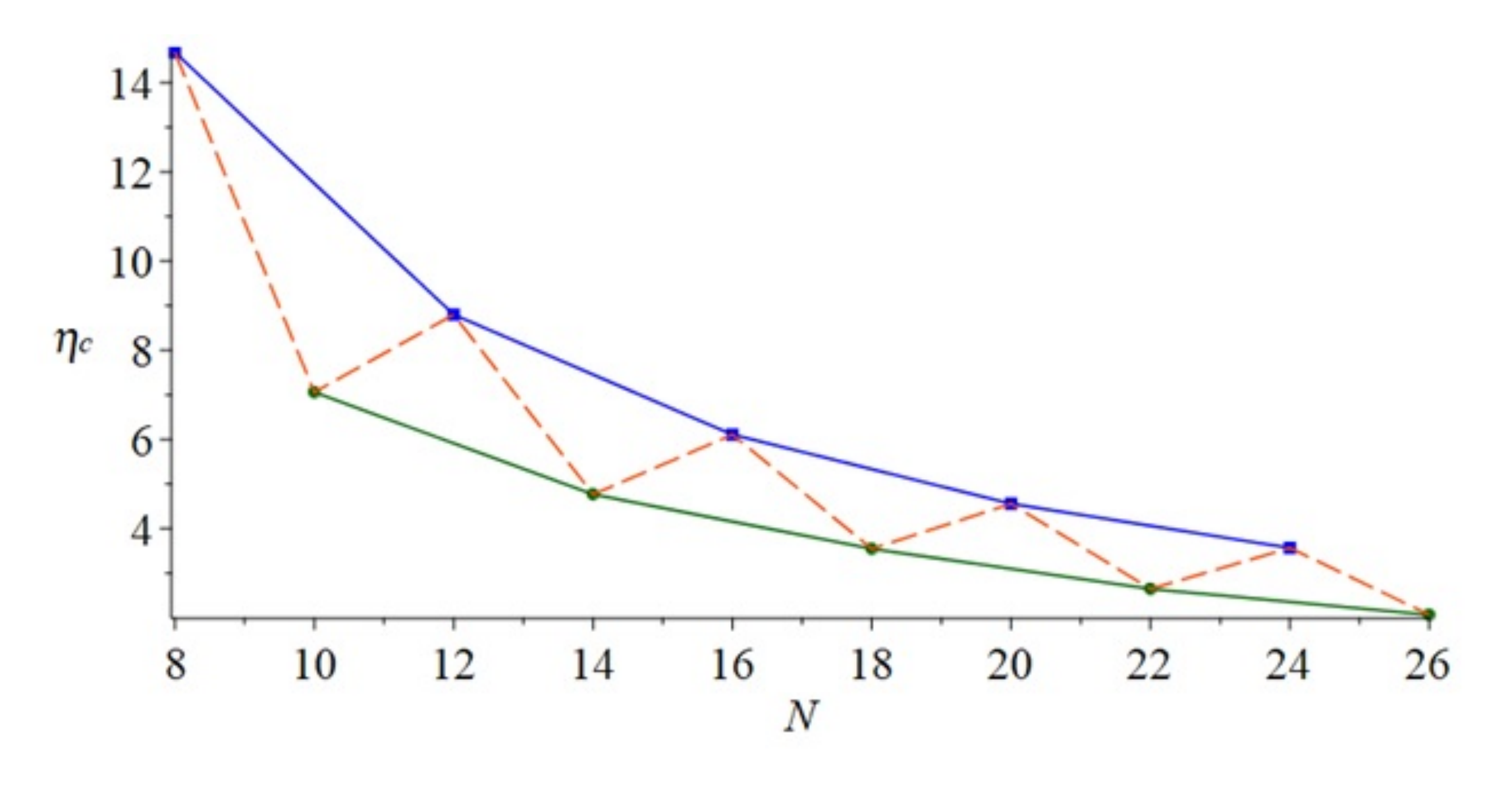} \caption{The parity effect for critical strain $\eta_{c}(N)$ for bi-structures with one long bond in the carbon chains. Blue and green colors (squares and circles) are used for cases of even and odd $N/2$, respectively, while dotted orange curve connects the points of critical strain for all values $N$ of the considered chains (color online).} \label{fig5}
\end{figure}

The dependence $\eta_{c}(N)$ on the total number $N$ of atoms in the carbon chain is represented by a zigzag dashed curve (see Fig.~\ref{fig5}). 

However, if we depict the dependence $\eta_{c}(N)$ {\it separately} for even and odd values of $N/2$, these dependences turn out to be {\it monotonic}, which can be approximated by low-degree polynomials. This fact is demonstrated in Fig.~\ref{fig5}, where blue plot corresponds to even number $N/2$, while green plot corresponds to odd number $N/2$.

\subsection{Coexistence of two energy minima for the strained carbon chains \label{Seq5.2}}

As was already mentioned in Sec.~\ref{Seq1.2}, there are works (see,~\cite{12,13,14,15,16,17,18}) devoted to study the static structures of finite carbon chains that are close to those of cumulene or polyyne in infinite chains. As to our understanding, the authors of these papers choose the equidistant atomic configuration of the strained carbon chain (cumulene structure) as the initial approximation and then refine it with the aid of the DFT-simulation. Such refining is usually performed with the aid of some descent method (as a rule this is the method of conjugate gradients), which leads to some {\it local minimum} of the system potential energy. Hereafter we refer to this minimum as min1.

On the other hand, our bi-structures, arising above the critical value $\eta_{c}(N)$ of the strain, correspond to another minimum of the chain energy, which we will call min2, and to get to this minimum we proceed from another initial approximation. It should be emphasized that both minima {\it coexist} in the system simultaneously, and the choice of one of them by the descent method depends only on the corresponding initial approximation.

The distance between minima min1 and min2 in multidimensional configuration space of all atoms of the carbon chain increases rapidly with increasing of the strain $\eta$.

It is interesting to consider the energy profiles of the system along the straight line connecting the two above-mentioned minima. Let the coordinates of min1 and min2 be $(x_{1}, x_{2}, ..., x_{N})$ and $(y_{1}, y_{2}, ..., y_{N})$, respectively, while the current point along the line in the N-dimensional space has coordinates $(z_{1}, z_{2} , ..., z_{N})$. If this point really lies on a line passing through our two minima, its coordinates must satisfy the equation $z_{i}=x_{i}+k\cdot (y_{i}-x_{i}), \,i=1..N$, where $k$ is the current parameter. If $k=0$, then $z_{i}=x_{i}$ for all $i$ and we obtain min1. In the case $k=1$ we obtain min2 since $z_{i}=y_{i}$.

\begin{figure}[h!] \centering
	\includegraphics[width=1\linewidth]{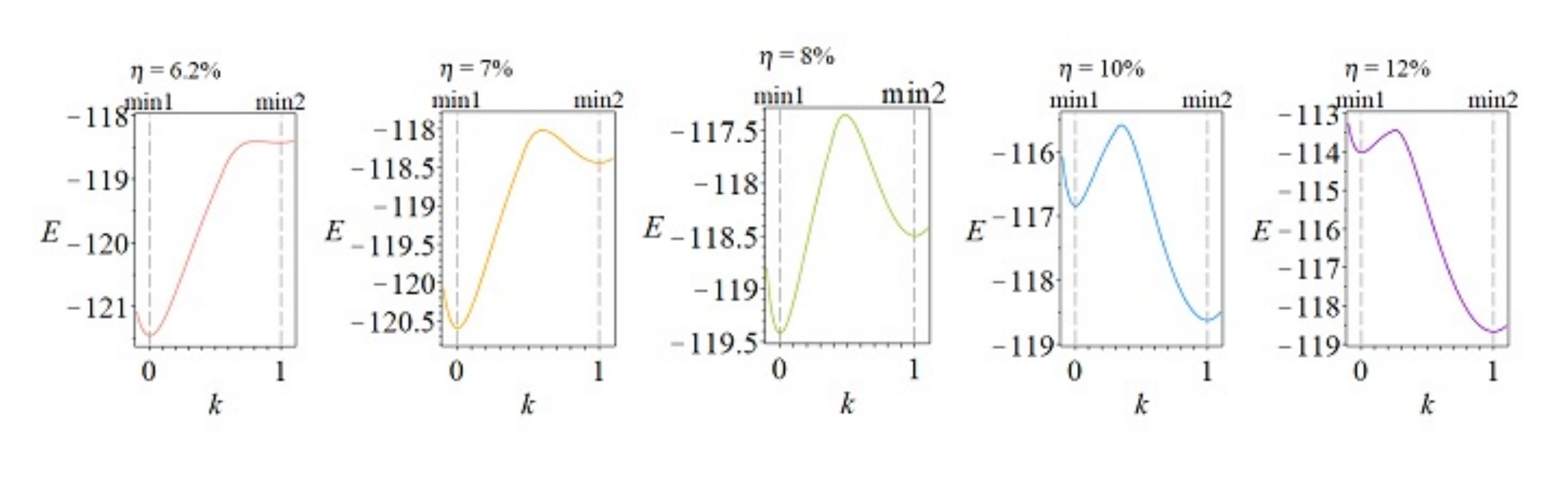} \caption{Energy profiles $E(k)$ for the carbon chain with $N=16$ atoms along the straight line between the two above-discussed minima for the strain values in the interval [$6\,{\%}$; $12\,{\%}$]. } \label{fig6}
\end{figure}

Figure~\ref{fig6} demonstrates energy profiles along the above line for different values of the strain. From these figures one can trace the change in the depth of the considered minima.

\subsection{Change in conductivity of the strained carbon chains when a bi-structure arises \label{Seq5.3}}

In Ref.~\cite{15, 19, 21}, the change in the conductivity of the infinite strained carbon chains due to the Peierls phase transition is discussed. This change occurs because of the Peierls distortion, corresponding to doubling of the primitive cell, an energy gap appears in the electron spectrum of cumulene, and the width of this gap increases with increasing the strain of the chain. As a result, the cumulene, which is a good conductor, transforms into polyyne, that is semiconductor or insulator, depending on the width of the gap.

A more significant change in electrical properties of carbon chains should occur in the vicinity of the bifurcations, which lead to the bi-structures. This phenomenon becomes obvious if we consider the situation within the framework of the strong coupling approach of the crystal theory. Indeed, the probability of electron hopping from one atom to another is determined by the degree of the overlapping of wave functions of neighboring atoms. On the other hand, such overlapping substantially depends on the distance between atoms, which increases dramatically when the long bond appears at the center of our bi-structures.

Calculation of the electron density between atoms of the carbon chain by DFT modeling shows that it decreases by several orders of magnitude in the region of the long bond, and this makes it difficult to depict it against the electron density in the region of short bonds. Nevertheless, in the next section, we show the electron density change during bifurcation, which leads to the bi-structure with {\it two} long bonds (see Fig.~\ref{fig12}). Note that despite the practical absence of the atomic orbitals overlap, attraction of two subchains in the vicinity of the long bond is provided by the van der Waals forces.

\subsection{Bi-structures with one long bond in the chains of boron atoms \label{Seq5.4}}

As was already noted, bi-structures found in the Lennard-Jones chains should be sufficiently universal objects, i.e. they can exist in any monoatomic chain under reasonable physical assumptions about the form of interatomic interactions. We have tested this hypothesis with the aid of DFT modeling of carbon and boron chains. Bi-structures with one long bond in carbon chains were considered in the previous sections of this work, and now we will carry out a similar analysis for the boron chains.

\begin{figure}[h!] \centering
	\includegraphics[width=1\linewidth]{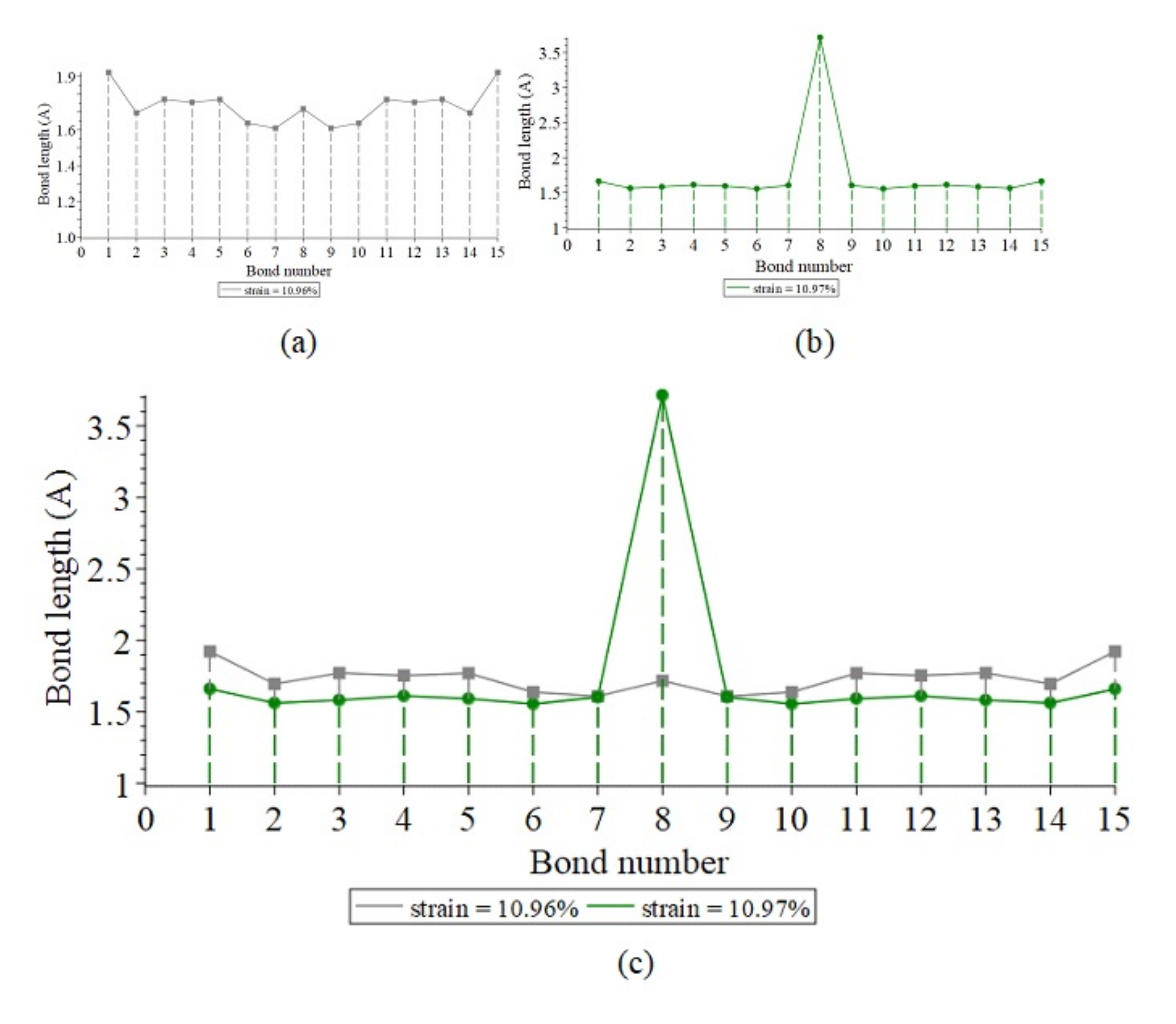} \caption{The bond lengths (BLs) for the chain of $N=16$ boron atoms: (a) before and (b) after bifurcation. The same bonds depicted in one picture (c) (color online).} \label{fig7}
\end{figure}

In Fig.~\ref{fig7}, which is similar to Fig.~\ref{fig4} for the carbon chain, we show the bi-structure in the boron chain directly before and after the critical bifurcation, while the corresponding numerical information is given in Table~\ref{t6}.

\begin{table}\centering
	\caption{Bond lengths before and after bifurcation for the chain of $N=16$ boron atoms, corresponding to the bi-structure with one long bond.}
	\renewcommand{\tabcolsep}{0.02\linewidth}
	\label{t6}

	\begin{tabular}{ccc}
		\hline
		 \specialcell{Number   of \\bond length} & \specialcell{$\eta=10.96 \,{\%}$ \\before bifurcation} & \specialcell{$\eta=10.97 \,{\%}$ \\after 
bifurcation} \\  

\hline

1	    & 1.923	& 1.660 \\
\hline
2	    & 1.696	& 1.562 \\
\hline
3	    & 1.772	& 1.582 \\
\hline
4	    & 1.755	& 1.611 \\
\hline
5	    & 1.771	& 1.592 \\
\hline
6	    & 1.638	& 1.555 \\
\hline
7	    & 1.609	& 1.603 \\
\hline
\rowcolor{lightgray} 8	    & 1.718	& 3.714 \\
\hline
9	    & 1.609	& 1.603 \\
\hline
10	    & 1.638	& 1.555 \\
\hline
11	    & 1.771	& 1.592 \\
\hline
12	    & 1.755	& 1.611 \\
\hline
13	    & 1.772	& 1.582 \\
\hline
14	    & 1.696	& 1.562 \\
\hline
15	    & 1.923	& 1.660 \\
\hline \hline
$E/N, eV$	& -4.24	& -4.16 \\

		\hline
	\end{tabular} 
\end{table}

Values of the critical strain for boron chains with different number of atoms and bond lengths after bifurcation, leading to the bi-structure with one long bond, are presented in Table~\ref{t7}.

\begin{table}\centering
	\caption{Parameters of the bi-structures with one long bond for boron chains with different number $N$ of atoms obtained by DFT modeling.}
	\renewcommand{\tabcolsep}{0.02\linewidth}
	\label{t7}

{\rowcolors{4}{lightgray}{white} 
	\begin{tabular}{ccccc}
		\hline
		 \multirow{2}*{$N$} & \multirow{2}*{Critical strain $\eta_{c}, \,{\%}$} & \multirow{2}*{Long bond, $\buildrel_\circ \over {\mathrm{A}}$} & \multicolumn{2}{c}{Short bonds, $\buildrel_\circ \over {\mathrm{A}}$} \\
		 \cline{4-5}
		 & & & Min & Max \\
		\hline
8 	&	18.91	&		3.185	&		1.574	&		1.714 \\ \hline
10	&	17.05	&		3.454	&		1.551	&		1.754 \\ \hline
12	&	14.61	&		3.717	&		1.566	&		1.668 \\ \hline
14	&	12.91	&		3.939	&		1.551	&		1.637 \\ \hline
16	&	10.97	&		3.714	&		1.555	&		1.660 \\ \hline
18	&	9.91 	&		3.885	&		1.546	&		1.661 \\ \hline
20	&	8.91 	&		3.793	&		1.554	&		1.658 \\ \hline
22	&	8.21 	&		3.914	&		1.548	&		1.644 \\ \hline
24	&	7.48 	&		3.857	&	 	1.548	&	 	1.646 \\ \hline
26	&	6.71 	&		3.787	&		1.547	&		1.656 \\		

		\hline
	\end{tabular} }
\end{table}

It is interesting that unlike carbon chains, the parity effect for the individual subchains of the boron chain is barely noticeable, but one can consider the parity effect for the whole chain of boron atoms.

\begin{figure}[h!] \centering
	\includegraphics[width=1\linewidth]{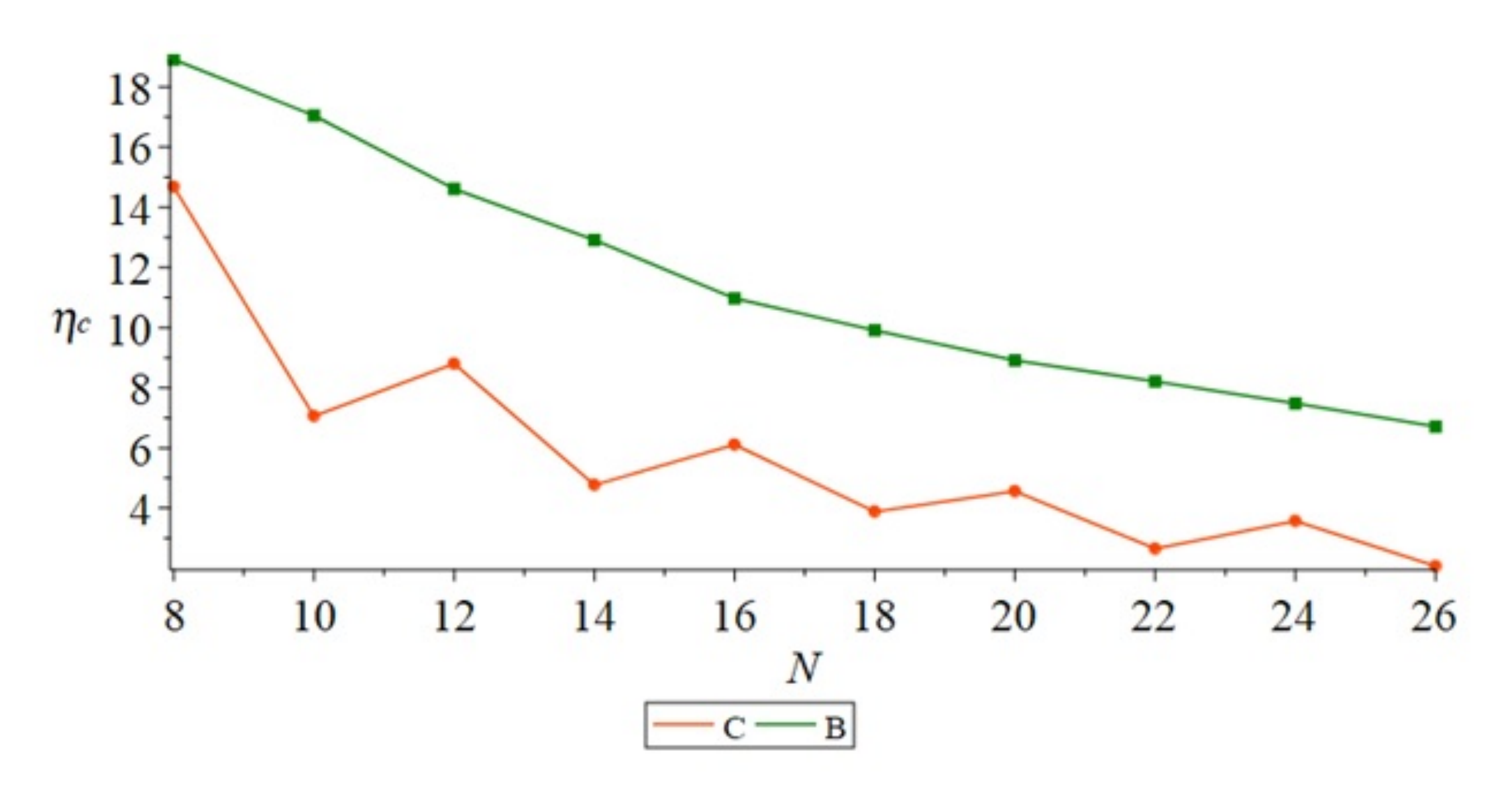} \caption{Comparison of the parity effect for carbon and boron chains (color online).} \label{fig8}
\end{figure}

This fact is demonstrated by Fig.~\ref{fig8}.

\section{Bi-structures with two long bonds in the strained monoatomic chains \label{Seq6}}

In the previous sections, we discussed the appearance of bi-structures with one long bond in strained carbon and boron chains and their properties. The symmetry of such bi-structure is determined by the inversion located between two central atoms of the chain, i.e. at the middle of the long bond of the bi-structure. Here we show that a different type of bi-structures can exist in the strained carbon chains, namely, the bi-structures with two long bonds and inversion located at the central atom of the chain. To study them, we apply the same method that was used for studying bi-structures with one long bond. Firstly, we prove that {\it exact homogeneous} bi-structures of this type can exist in the L-J chains. Then we show that these bi-structures can be revealed in the carbon chains with the aid of DFT modeling, if we use the adequate initial configuration, which is suggested by the study of L-J chains, and show that such bi-structures are {\it heterogeneous}.

\subsection{Bi-structures with two long bonds in the strained L-J chains \label{Seq6.1}}

\begin{figure}[h!] \centering
	\includegraphics[width=1\linewidth]{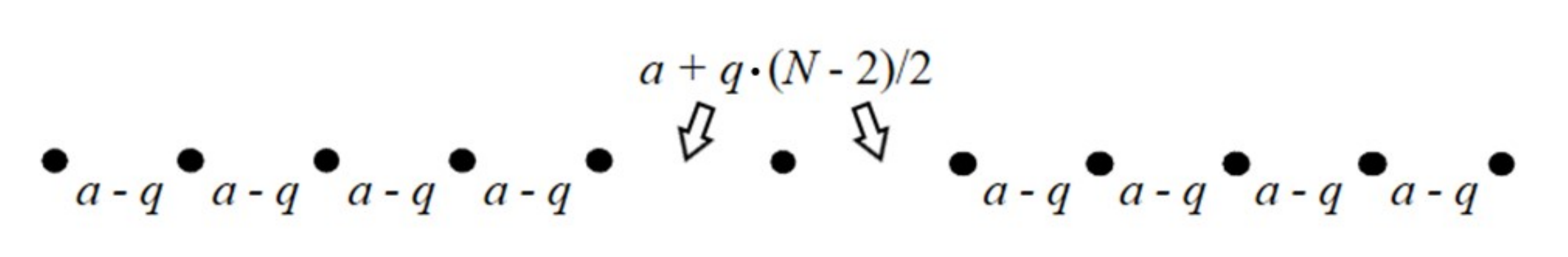} \caption{Bi-structure with two long bonds in the strained L-J chain with $N=11$ atoms.} \label{fig9}
\end{figure}

Bi-structures with two long bonds can exist in the chain with {\it odd} number of atoms $N$ with inversion located at the center of this chain. In Fig.~\ref{fig9}, we show such bi-structure in the chain with $N=11$ atoms. The inversion is located at the central atom, and the bi-structure is formed by two identical {\it subchains} located symmetrically with respect to this atom.

Let us show that such a structure is indeed possible in the strained L-J chain.

Similar to the case of the bi-structure with one long bond (bi-structure 1), all atoms of the chain located at the junction of {\it two short bonds} are automatically in equilibrium, because these bonds are {\it identical}. Therefore, it is enough to require the equilibrium state of atoms located only at the junction of the short and long bonds. This condition leads to the following force equation:

\begin{equation}
\label{eq4}
f(a-q)=f\left [  a+\frac{q\cdot(N-2)}{2}\right ]
\end{equation}

Here $a$ is the interatomic distance of the strained cumulene, while arguments of the L-J force on l.h.s. and on r.h.s. of this equation are the short bond $(a-q)$ and the long bond $a+q\cdot(N-2)/2$, respectively. 

Obviously, Eq.~\ref{eq4} is only slightly different from the force equation for the case of the bi-structure with one long bond (see Eq.~\ref{eq3}). We must solve Eq.~\ref{eq4} for an unknown $q$, which shows how much is the short length less than $a$. 

Similar to the force equation~\ref{eq3}, a rigid bifurcation appears with increasing of the chain strain, and it corresponds to the stable static bi-structure with two long bonds. The parameters of this bi-structure are given in Table~\ref{t8}.

\begin{table}\centering
	\caption{The critical strain $\eta_{c}(N)$ and bond lengths of the appearing homogeneous bi-structure with two long bonds as the functions of the number $N$ of atoms in the L-J chains.}
	\renewcommand{\tabcolsep}{0.02\linewidth}
	\label{t8}
	{\rowcolors{2}{lightgray}{white} 
	\begin{tabular}{cccc}
		\hline
		$N$ & Critical strain $\eta_{c},\,{\%}$ & Long bond & Short bond \\
		\hline
7	&	10.05	&	1.316	&	1.195 \\ \hline
9	&	8.97	&	1.369	&	1.175 \\ \hline
11	&	8.05	&	1.404	&	1.175 \\ \hline
13	&	7.31	&	1.439	&	1.158 \\ \hline
15	&	6.70	&	1.463	&	1.154 \\ \hline
17	&	6.20	&	1.494	&	1.149 \\ \hline
19	&	5.78	&	1.522	&	1.146 \\ \hline
21	&	5.41	&	1.534	&	1.144 \\ \hline
23	&	5.10	&	1.560	&	1.142 \\ \hline
25	&	4.82	&	1.569	&	1.141 \\ \hline
27	&	4.58	&	1.591	&	1.139 \\ \hline
33	&	4.00	&	1.642	&	1.136 \\ \hline
43	&	3.32	&	1.683	&	1.134 \\ \hline
53	&	2.87	&	1.749	&	1.131 \\ \hline
103	&	1.77	&	1.905	&	1.127 \\
	    \hline	
	\end{tabular}}
\end{table}

\begin{figure}[h!] \centering
	\includegraphics[width=1\linewidth]{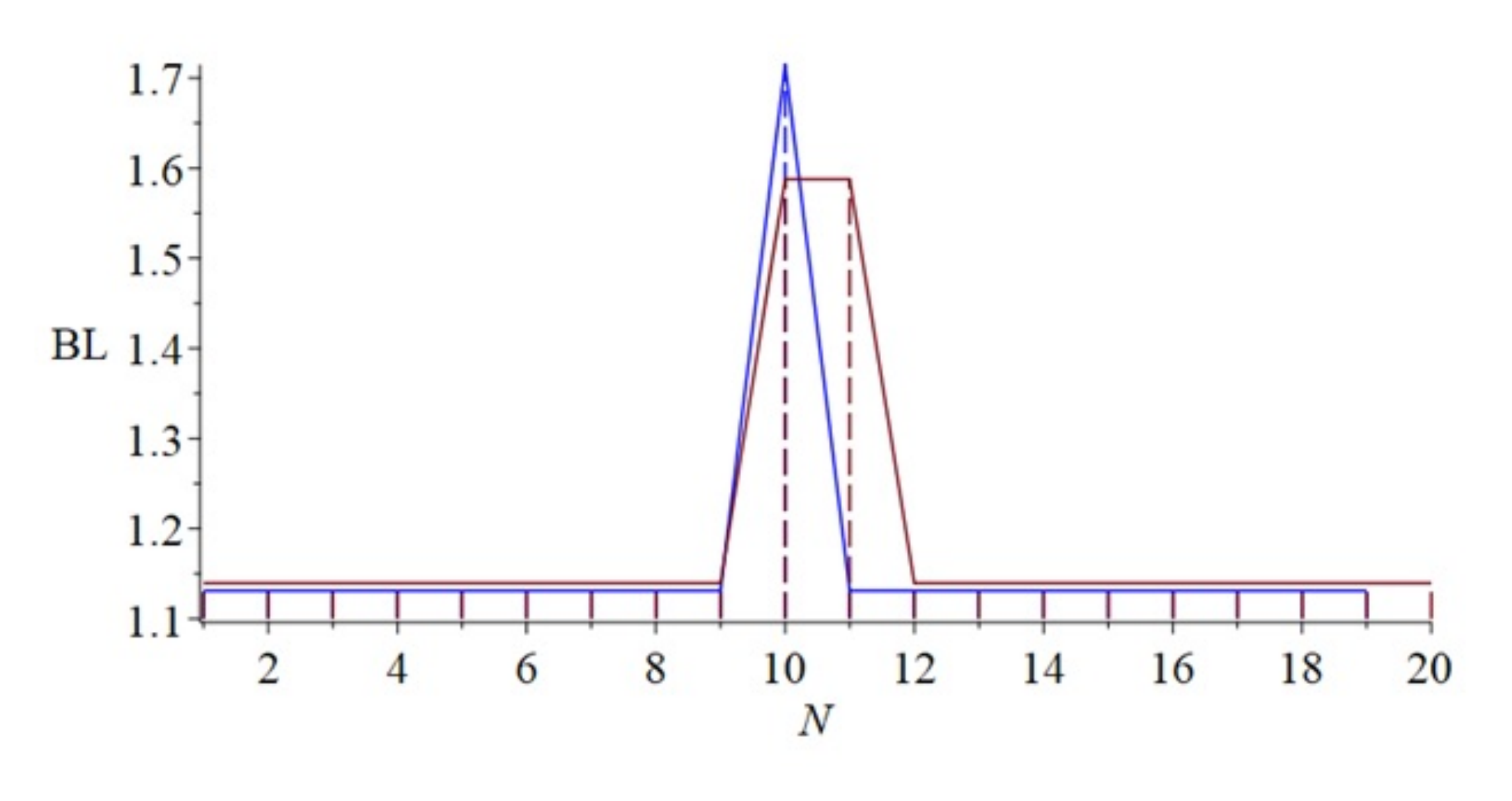} \caption{The bond lengths (BL) for bi-structure 1 (blue color) and those for bi-structure 2 (red color) in the L-J chain with $3.5\,{\%}$ and $5.5\,{\%}$ of the strain, respectively (color online).} \label{fig10}
\end{figure}

This Table shows that for the bi-structure with two long bonds (bi-structure 2) the critical strain, corresponding to its appearance, tends to decrease with increasing the number $N$ of particles in L-J chain, as well as for the bi-structure with one long bond (bi-structure 1) discussed in Sec.~\ref{Seq3}. Comparison of bond lengths in the bi-structure 1 ($N$ is even) and in the bi-structure 2 ($N$ is odd) at close values $N$ shows that both long and short bonds in the latter bi-structure are larger than those in the former bi-structure. In Fig.~\ref{fig10}, this difference is demonstrated by the example of the chains with $N=20$ (for bi-structure 1) and $N=21$ (for bi-structure 2).

\subsection{Bi-structures with two long bonds in the strained carbon chains obtained with the aid of DFT modeling \label{Seq6.2}}

It is necessary to choose an appropriate initial approximation to find bi-structures with two long bonds in the strained carbon chains within the framework of the DFT theory. From Fig.~\ref{fig9}, corresponding to such bi-structure in the strained L-J chain, it is clear that one can take as initial configuration two identical cumulene subchains equally spaced left and right from the central atom of the chain, at which the inversion is located. Refining of this initial structure with the aid of DFT modeling leads to the following {\it heterogeneous} bi-structures presented in Table~\ref{t9}, which are similar to homogeneous bi-structures in the strained L-J chains.

\begin{table}\centering
	\caption{Parameters of bi-structures with two long bonds in strained carbon chains for different $N$, calculated with the aid of DFT modeling.}
	\renewcommand{\tabcolsep}{0.02\linewidth}
	\label{t9}

{\rowcolors{4}{lightgray}{white} 
	\begin{tabular}{ccccc}
		\hline
		 \multirow{2}*{$N$} & \multirow{2}*{Critical strain $\eta_{c}, \,{\%}$} & \multirow{2}*{Long bond, $\buildrel_\circ \over {\mathrm{A}}$} & \multicolumn{2}{c}{Short bonds, $\buildrel_\circ \over {\mathrm{A}}$} \\
		 \cline{4-5}
		 & & & Min & Max \\
		\hline
9	&	16.75 	&	2.030	&	1.324	&	1.368 \\ \hline
11	&	10.84	&	2.021	&	1.292	&	1.310 \\ \hline
13	&	10.96	&	2.060	&	1.296	&	1.337 \\ \hline
15	&	7.80	&	2.059	&	1.280	&	1.307 \\ \hline
17	&	8.10	&	2.074	&	1.290	&	1.331 \\ \hline
19	&	6.01	&	2.069	&	1.278	&	1.311 \\ \hline
21	&	6.37	&	2.092	&	1.286	&	1.327 \\ \hline
23	&	4.82	&	2.073	&	1.277	&	1.313 \\ \hline
25	&	5.20	&	2.100	&	1.283	&	1.326 \\ \hline
27	&	4.03	&	2.079	&	1.277	&	1.315 \\ \hline
	\end{tabular} }
\end{table}

It should be noted that the largest discrepancies in the values of the critical strain calculated in the framework of DFT modeling and in the Lennard-Jones model take place for a small number of atoms ($N<19$), while for a larger number of atoms in the chain these discrepancies are within $1\,{\%}$ (see Fig.~\ref{fig13}).

\begin{figure}[h!] \centering
	\includegraphics[width=1\linewidth]{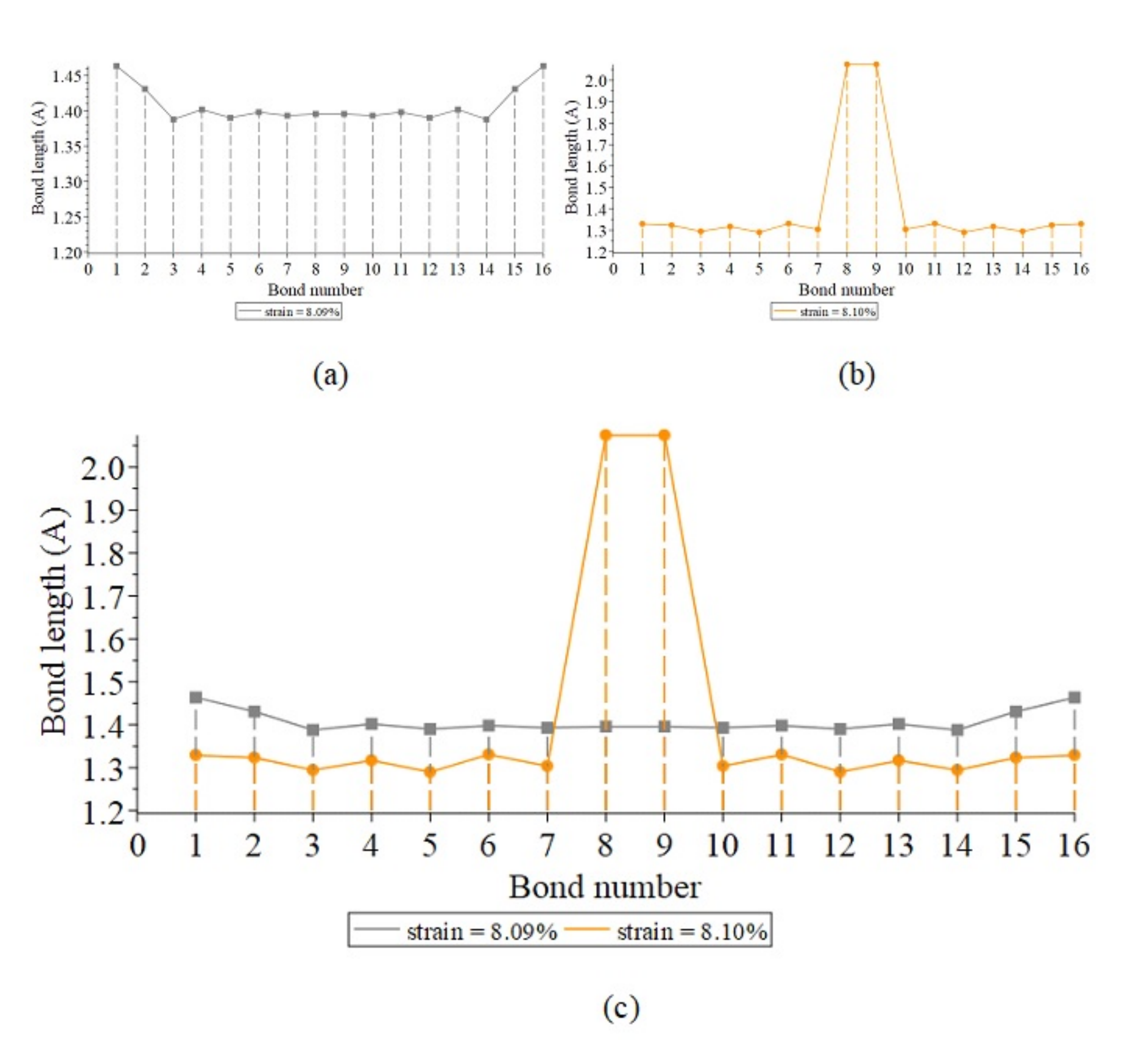} \caption{Bond lengths before (a) and after (b) bifurcation for the carbon chain with $N=17$ atoms. (c) the data of (a) and (b) combined in the same graph (color online).} \label{fig11}
\end{figure}

The bond lengths for the carbon chain with $N=17$ atoms before and after bifurcation are presented in Table~\ref{t10} and in Fig.~\ref{fig11}.

\begin{table}\centering
	\caption{Appearance of the bi-structure with two long bonds in the carbon chain with $N=17$ atoms. Bond lengths ($\buildrel_\circ \over {\mathrm{A}}$) before and after bifurcation obtained with the aid of DFT modeling.}
	\renewcommand{\tabcolsep}{0.02\linewidth}
	\label{t10}

	\begin{tabular}{ccc}
		\hline
		 Bond number & \specialcell{Bond length for \\ $\eta=8.09\,{\%}$ \\ (before bifurcation)} & \specialcell{Bond length for \\ $\eta=8.10\,{\%}$ \\ (after bifurcation)} \\
		\hline
1  	& 1.463 & 	1.329 \\ \hline
2 	& 1.431	& 	1.323 \\ \hline
3	& 1.388	& 	1.294 \\ \hline
4	& 1.402	& 	1.317 \\ \hline
5	& 1.390	& 	1.290 \\ \hline
6	& 1.398	& 	1.331 \\ \hline
7	& 1.393	& 	1.304 \\ \hline
\rowcolor{lightgray} 8	& 1.396	& 	2.074 \\ \hline
\rowcolor{lightgray} 9	& 1.396	& 	2.074 \\ \hline
10	& 1.393	& 	1.304 \\ \hline
11	& 1.398	& 	1.331 \\ \hline
12	& 1.390	& 	1.290 \\ \hline
13	& 1.402	& 	1.317 \\ \hline
14	& 1.388	& 	1.294 \\ \hline
15	& 1.431	& 	1.323 \\ \hline
16	& 1.463	& 	1.329 \\

\hline
\hline                   
$E/N, eV$ & -7.49	& -7.23 \\
		\hline
	\end{tabular}
\end{table}

\begin{figure}[h!] \centering
	\includegraphics[width=1\linewidth]{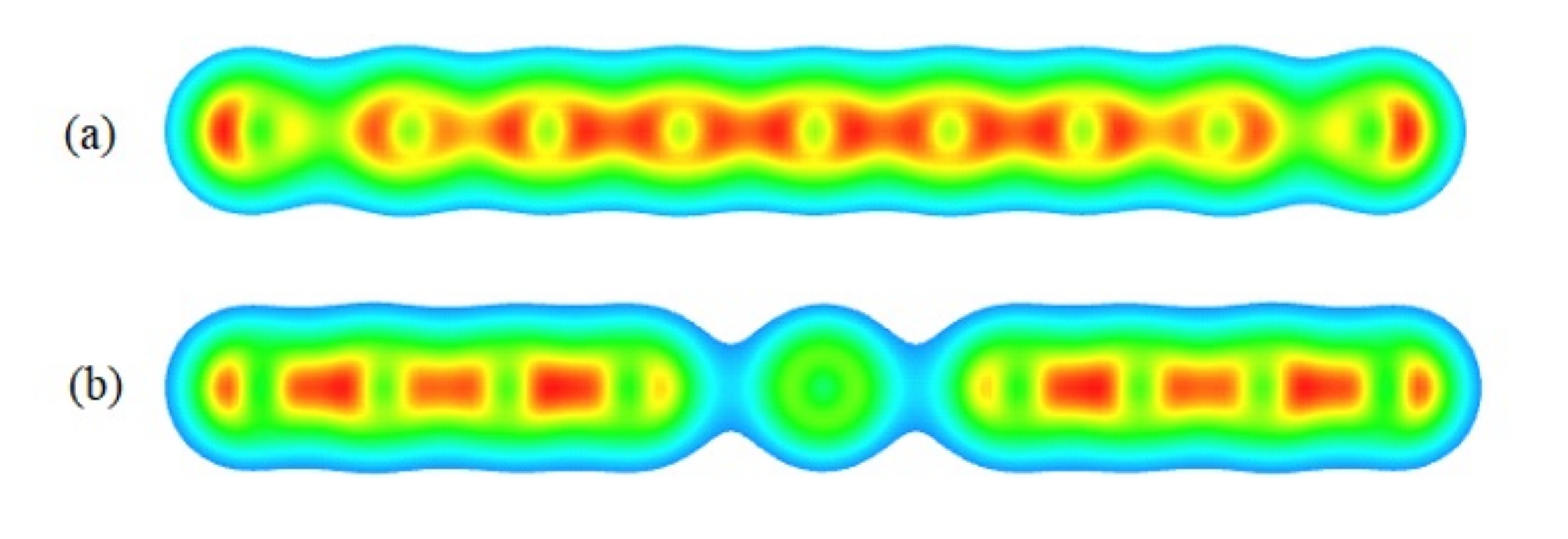} \caption{The electron density change in the carbon chain with $N=9$ atoms for bifurcation leading to the bi-structure with two long bonds: (a) for the strain $\eta=16.7\,{\%}$ (before bifurcation) and (b) for the strain $\eta=17.0\,{\%}$ (after bifurcation) (color online).} \label{fig12}
\end{figure}

\begin{figure}[h!] \centering
	\includegraphics[width=1\linewidth]{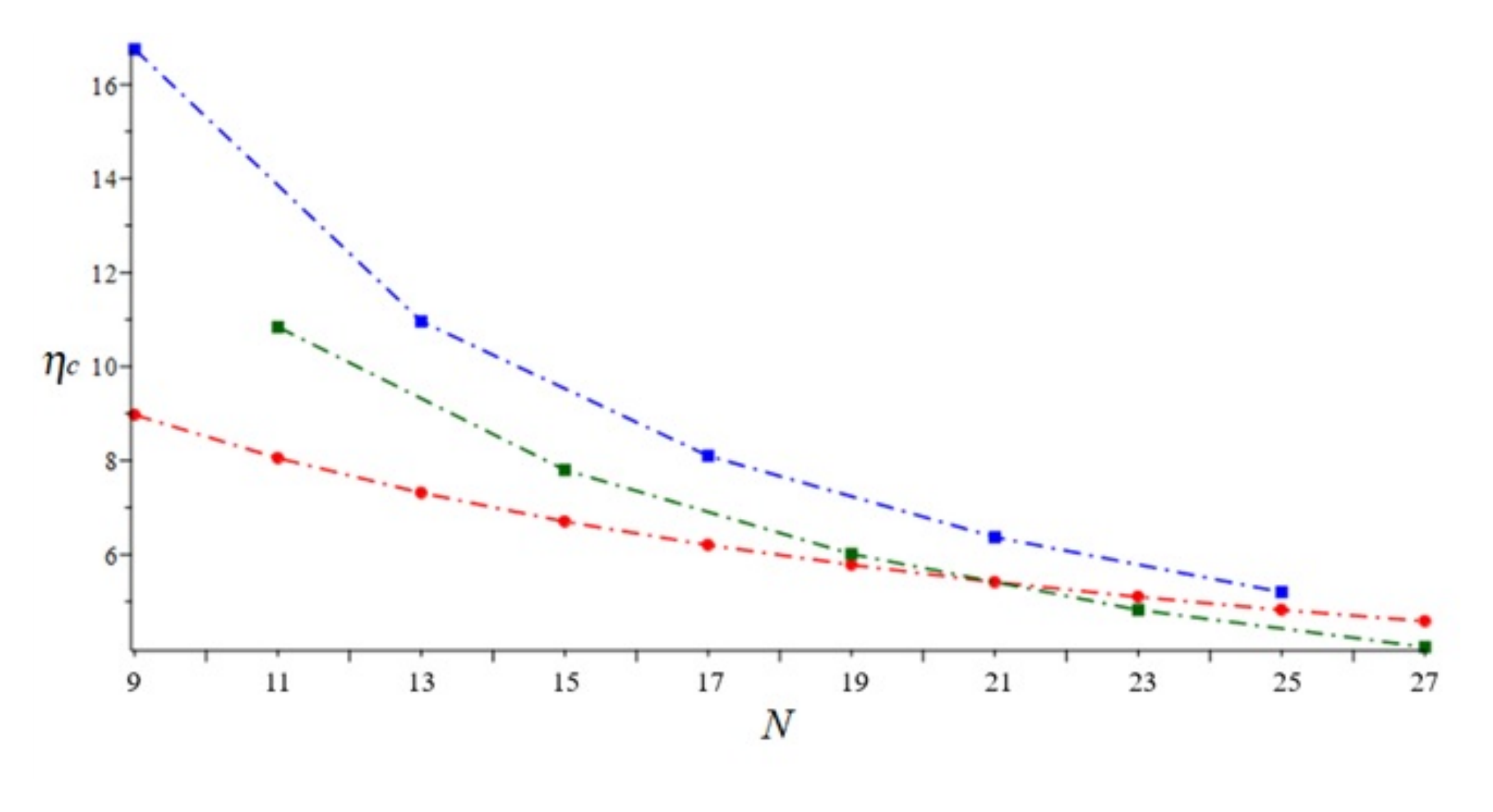} \caption{Comparison of critical strain for appearing bi-structures with two long bonds in carbon chains calculated in the L-J model (red circles) and in the DFT model (blue and green squares) (color online).} \label{fig13}
\end{figure}

The appearance of long bonds in the central part of the carbon chain significantly changes the electron density distribution along the chain, which is illustrated in Figure~\ref{fig12}.

It is appropriate to comment on the comparison of the strain $\eta$ of chains in the L-J and DFT models (see Fig.~\ref{fig13}), because we deal with homogeneous structures in the first case and with heterogeneous ones in the second case. In both cases, we assume $\eta=(L_{1}-L_{0})/L_{0}$ where $L_{1}$ and $L_{0}$ are lengths of the chain with and without strain, respectively. This definition of the strain coincides with that used earlier for homogeneous structures $\eta=(a-a_{0})/a_{0}$, where $a$ and $a_{0}$ are sizes of the primitive cells of the chain in the presence and absence of the strain, respectively.

In Fig.~\ref{fig13}, we present the critical strain for appearing bi-structures with two long bonds in carbon chains calculated in the framework of the L-J and DFT-model. It can be seen from these results that the parity effect, discussed in Sec.~\ref{Seq5} for bi-structures with one long bond also takes place for the bi-structures with two long bonds in the carbon chains.

With the aid of DFT modeling, we examined two types of static bi-structures in carbon chains that arise on the assumption of different localization of inversion, which can be localized between two central atoms (bi-structure with one long bond) or at the central atom (bi-structure with two long bonds). The first type of these bi-structures occurs in chains with an even number $N$ of atoms, and the second type can exist in chains with an odd number $N$. The above mentioned structures correspond to different energy minima of the carbon chains, and their depth determines the energy $E(eV)$ per atom of the chain. These values of energy for different $N$ are listed in Table~\ref{t11}. 

\begin{table}\centering
	\caption{Energy per one atom of the carbon chain for bi-structures obtained in the framework of the DFT modeling for the critical strains.}
	\renewcommand{\tabcolsep}{0.02\linewidth}
	\label{t11}

	\begin{tabular}{cccc}
		\hline
		 \multirow{3}*{$N$} & Bi-stucture with  &  \multirow{3}*{$N$} & Bi-stucture with  \\
		 & one long bond & & two long bonds  \\
		 & $E${\it , eV} & & $E${\it , eV}  \\
		\hline
8	&	-6.44	&	9	&	-6.26 \\ \hline
10	&	-7.05	&	11	&	-6.74 \\ \hline
12	&	-7.09	&	13	&	-6.90 \\ \hline
14	&	-7.35	&	15	&	-7.14 \\ \hline
16	&	-7.40	&	17	&	-7.23 \\ \hline
18	&	-7.56	&	19	&	-7.37 \\ \hline
20	&	-7.58	&	21	&	-7.43 \\ \hline
22	&	-7.69	&	23	&	-7.53 \\ \hline
24	&	-7.70	&	25	&	-7.56 \\ \hline
26	&	-7.77	&	27	&	-7.63 \\ \hline
	\end{tabular} 
\end{table}

It can be seen from the data presented in Table~\ref{t11}, that the bi-structures with one long bond are slightly more stable than those with two long bonds because the energies, at close values of $N$, for the former bi-structures correspond to deeper minima than for the latter bi-structures.

\subsection{Bi-structure with two long bonds in the boron chains \label{Seq6.3}}

Similar to the carbon chains, the bi-structures 2 in the DFT model of boron chains, appear {\it abruptly} with increasing of the strain, i.e. as a result of a {\it rigid bifurcation}. As an example, we consider the appearance of the bi-structure 2 in the boron chain with $N=17$ atoms. Table~\ref{t12} shows the lengths of the interatomic bonds directly {\it before} and {\it after} the bifurcation, i.e. after passing through the critical value of the strain $\eta_{c}(N)=13.16\,{\%}$. The first column contains the numbers of interatomic bonds for $i=1..16$, where $i$ is the bond number between atoms with numbers $i$ and $i+1$.

\begin{table}\centering
	\caption{Appearance of the bi-structure with two long bonds in the boron chain with $N=17$ atoms. Bond lengths ($\buildrel_\circ \over {\mathrm{A}}$) before and after bifurcation obtained with the aid of DFT modeling.}
	\renewcommand{\tabcolsep}{0.02\linewidth}
	\label{t12}

	\begin{tabular}{ccc}
		\hline
		 Bond number & \specialcell{Bond length for \\ $\eta=13.15\,{\%}$ \\ (before bifurcation)} & \specialcell{Bond length for \\ $\eta=13.16\,{\%}$ \\ (after bifurcation)} \\
		\hline
1  	& 2.333 & 	1.702 \\ \hline
2 	& 1.692	& 	1.580 \\ \hline
3	& 1.636	& 	1.596 \\ \hline
4	& 1.640	& 	1.629 \\ \hline
5	& 1.695	& 	1.614 \\ \hline
6	& 1.706	& 	1.575 \\ \hline
7	& 1.718	& 	1.616 \\ \hline
\rowcolor{lightgray} 8	& 1.744	& 	2.853 \\ \hline
\rowcolor{lightgray} 9	& 1.744	& 	2.853 \\ \hline
10	& 1.718	& 	1.616 \\ \hline
11	& 1.706	& 	1.575 \\ \hline
12	& 1.695	& 	1.614 \\ \hline
13	& 1.640	& 	1.629 \\ \hline
14	& 1.636	& 	1.596 \\ \hline
15	& 1.692	& 	1.580 \\ \hline
16	& 2.333	& 	1.702 \\

\hline
\hline                   
$E/N, eV$ & -4.23	& -4.03 \\
		\hline
	\end{tabular}
\end{table}

As in the case of carbon chains, the bi-structures 2 in the finite boron chains are heterogeneous and their parameters depend on the number of atoms in the chain. In Table~\ref{t13}, the values of the critical strain $\eta_{c}$ for boron chains with different number of atoms and bond lengths after bifurcation, leading to the bi-structure 2, are presented. Because of heterogeneity of the bi-structure 2, in this table we give the minimal and maximal values of the short bonds.

\begin{table}\centering
	\caption{Parameters of the bistructures 2 for the boron chains for different number of their atoms.}
	\renewcommand{\tabcolsep}{0.02\linewidth}
	\label{t13}

{\rowcolors{4}{lightgray}{white} 
	\begin{tabular}{ccccc}
		\hline
		 \multirow{2}*{$N$} & \multirow{2}*{Critical strain $\eta_{c}, \,{\%}$} & \multirow{2}*{Long bond, $\buildrel_\circ \over {\mathrm{A}}$} & \multicolumn{2}{c}{Short bonds, $\buildrel_\circ \over {\mathrm{A}}$} \\
		 \cline{4-5}
		 & & & Min & Max \\
		\hline
9	&	22.94	&	2.739	&	1.585	&	1.749 \\ \hline
11	&	19.32	&	2.810	&	1.560	&	1.746 \\ \hline
13	&	16.77	&	2.830	&	1.590	&	1.715 \\ \hline
15	&	14.78	&	2.841	&	1.585	&	1.704 \\ \hline
17	&	13.16	&	2.853	&	1.575	&	1.702 \\ \hline
19	&	11.72	&	2.863	&	1.569	&	1.709 \\ \hline
21	&	10.89	&	2.407	&	1.614	&	1.754 \\ \hline
23	&	10.17	&	2.832	&	1.578	&	1.708 \\ \hline
25	&	9.40	&	2.918	&	1.568	&	1.683 \\ \hline
27	&	8.74	&	2.863	&	1.572	&	1.701 \\ \hline
	\end{tabular} }
\end{table}

\begin{figure}[h!] \centering
	\includegraphics[width=1\linewidth]{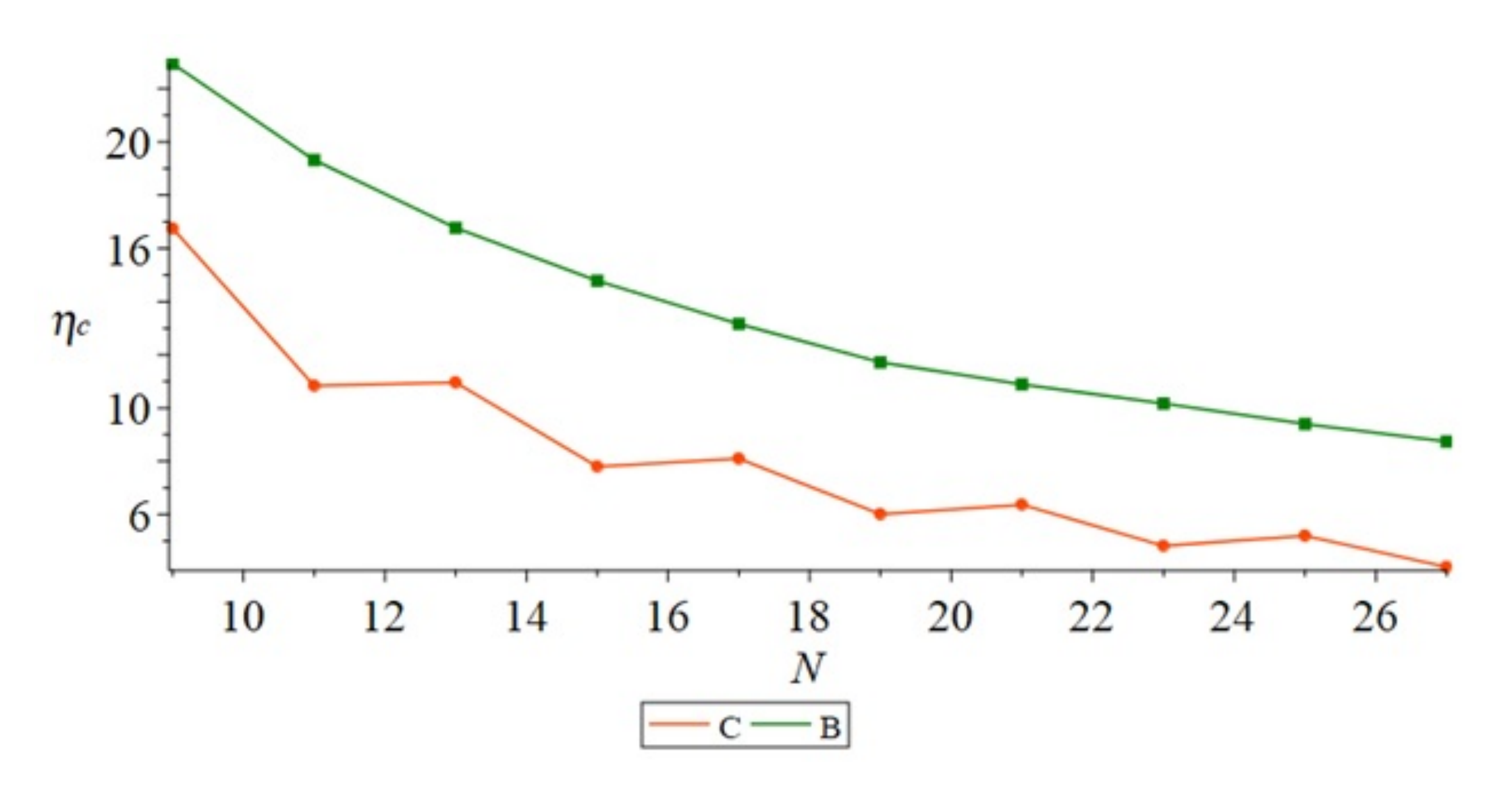} \caption{Parity law for the critical strain for the bi-structure 2 arising in the carbon and boron chains (color online).} \label{fig14}
\end{figure}

In Sec.~\ref{Seq5}, we have discussed the parity law for bi-structures 1 arising in the carbon and boron chains with different numbers of atoms. It is interesting to compare this law with that for bi-structures 2. Such comparison can be done by considering Fig.~\ref{fig14}.

The following properties of the critical strain are obvious from this figure.
\begin{enumerate} 
    \item For both carbon and boron chains the critical strain decreases with increasing the number $N$ of their atoms.
    \item For close $N$, the critical value for the boron chain is noticeably larger than that for the carbon chain. This property occurs for both types of bi-structures.
    \item The pronounced dependence of the critical strain on the {\it parity} of the number of atoms in the chain occurs only for carbon chains (see zigzag curve in Figs.~\ref{fig8} and ~\ref{fig14}). This conclusion is true for the bi-structure 1, as well as for the bi-structure 2.
\end{enumerate}

\begin{figure}[h!] \centering
	\includegraphics[width=1\linewidth]{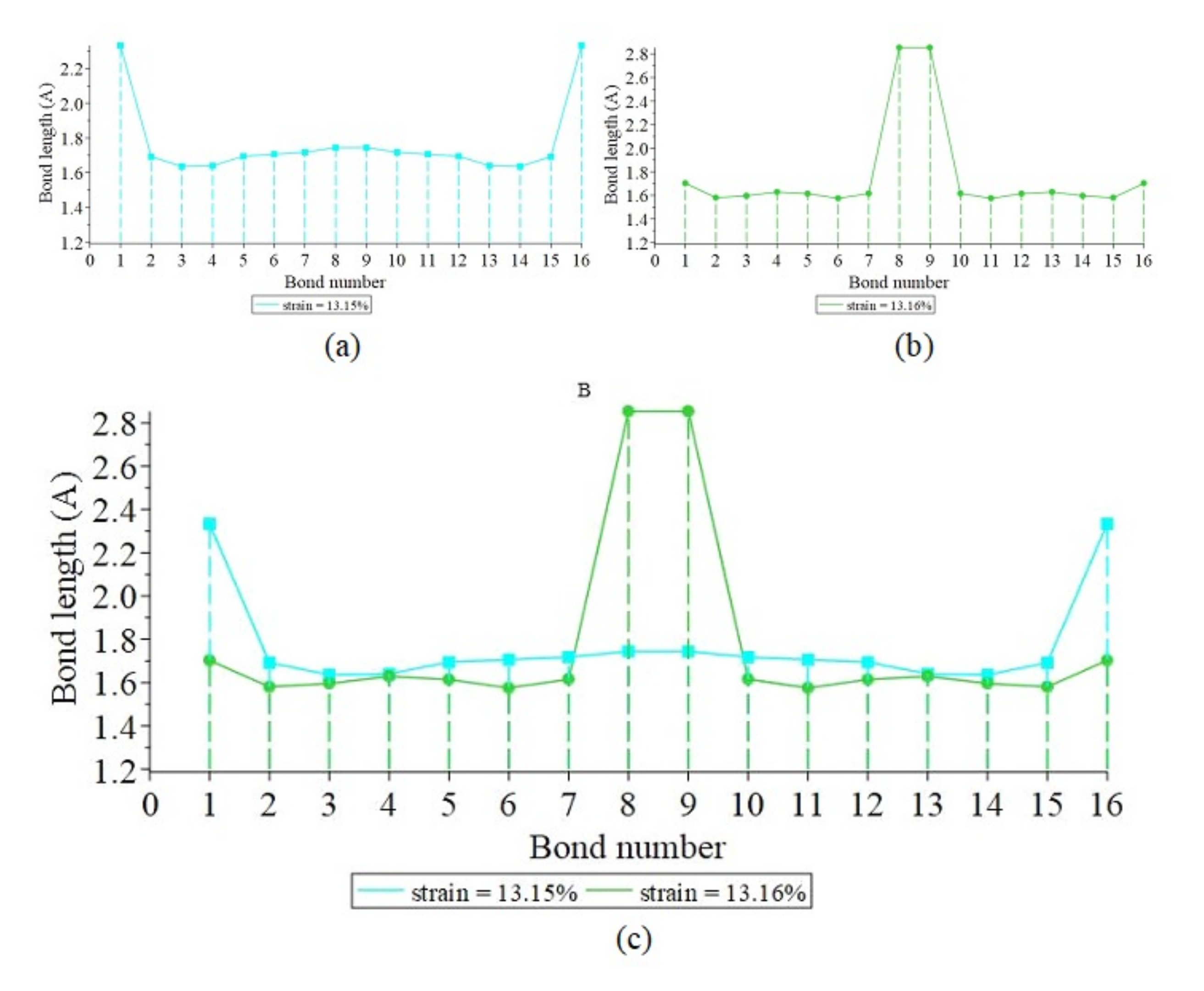} \caption{The structure of the boron chain with $N=17$ atoms before and after bifurcation, which leads to appearance of the bi-structure 2 (color online).} \label{fig15}
\end{figure}

The structure of the boron chain with $N=17$ atoms before and after bifurcation, leading to the appearance of the bi-structure 2, is shown in Fig.~\ref{fig15}. It can be seen from this figure that the longest of all short bonds occur for atoms located directly near the ends of the chains for both their types.

\section{Additional discussions and conclusion \label{Seq7}}

Two new types of static structures in the strained monoatomic chains of finite size are studied in the present paper. We describe the physical mechanism of their appearance by example of chains with Lennard-Jones potential (L-J model) and show that these structures are {\it exact} solutions of the corresponding static equations. We also present some arguments for the possibility of their existence in chains with arbitrary interatomic potentials. 

The use of the L-J model allows us to find the initial configurations of carbon and boron chains, which lead to the bi-structure 1 and bi-structure 2 as specific equilibrium states of these chains above some critical values of the strain. These static bi-structures are {\it homogeneous} in the L-J model, while they are {\it heterogeneous} in the DFT model of carbon and boron chains.

Bi-structure 1 (B-1) has one long bond and bi-structure 2 (B-2) has two long bonds in the central part of the chain. Other bonds in both bi-structures are short bonds. In the L-J chain, all short bonds are identical in each of the bi-structures, while in the bi-structures of carbon and boron chains they are of different size. Because of this fact we use the term ``homogeneous'' for bi-structures in L-J chains, and the term ``heterogeneous'' for those in the carbon and boron chains.

In fact, B-1 represents two identical subchains, which are separated by one long bond, while in B-2 such subchains are separated by two long bonds.

\subsection{Phase transitions and bifurcations of crystal structures associated with the spontaneous symmetry breaking \label{Seq7.1}}

Without going into the subtleties of the concept of phase transitions in crystals, we note that most of them are associated with a spontaneous symmetry breaking with changing of such scalar parameters as temperature and pressure~\cite{27}.

There exist phase transitions of the first order (the structure of a crystal changes abruptly) and transitions of the second order (the structure of the crystal changes continuously).

Both types of these phase transitions are discussed for infinite crystal systems. In the case of {\it finite} systems, they correspond to transformation of the structure of the system associated with a {\it rigid bifurcation} (analog of the first-order phase transition) or a {\it soft bifurcation} (analog of the second-order phase transition).

In introduction, we have already discussed the paper~\cite{15}, in which the Peierls phase transition in the infinite strained carbon chain was examined, as well as the associated change of electron spectrum and other properties of the chain (from another point of view this transition was studied in our work~\cite{21}). The Peierls transition from the cumulene leads to doubling of the period of this one-dimensional crystal, due to which its symmetry is halved. Indeed, the cumulene symmetry group contains translations for all multiples of the period of the carbon one-dimensional lattice and all inversions of two types. These are the inversions that pass through all the atoms of the chain and through the middles of all neighboring atoms. As a result of doubling of the lattice period, half of these symmetry elements disappear, which leads to the above reduction of the chain symmetry group.

In the present work, we study bifurcations which lead to the static structures of the finite strained monoatomic chains with only one symmetry element. This is the inversion located at the middle of two central atoms for chains with even $N$, or at a central atom for chains with odd $N$, where $N$ is the number of atoms in the chain.

In the former case, the bi-structure with one long bond (B-1) appears as a result of the bifurcation, while the bi-structure with two long bonds (B-2) appears  in the latter case.

It is important to emphasize that both these bifurcations are {\it hard}, whereas the bifurcation corresponding to the Peierls transition in the infinite chain is {\it soft}. This distinction seems to be significant for using carbon chains as material for future applications in nanodevices in the field of straintronics, since an {\it abrupt} change of material properties at the critical value of the strain can be more preferable than the change in a continuous manner.

\subsection{Comparison of the bi-structures with one long bond in the strained carbon and boron chains \label{Seq7.2}}

The critical strain $\eta_{c}$ corresponding to the appearance of the bi-structure 1 decreases quite rapidly with increasing $N$ for both carbon and boron chains (see Table~\ref{t5} and Table~\ref{t7}).

For example, for the carbon chain with $N=14$ atoms, $\eta_{c}=4.77\,{\%}$, while for that with $N=26$ atoms it is equal to $\eta_{c}=2.07\,{\%}$. Thus, the bi-structure 1 arises in carbon chains, for the case of large $N$, at rather small values of the strain (it certainly exists at greater strains as well, but its parameters change significantly with increasing $\eta$).

For boron chains with the same number of atoms, the critical strain at which bi-structure 1 appears is much larger. For example, for the boron chain with $N=26$ atoms the critical strain is $\eta_{c}=6.71\,{\%}$ in contrast to $\eta_{c}=2.07\,{\%}$ for the carbon chain with the same $N$ (let us note that the corresponding value $\eta_{c}$ for the L-J chain with $N=26$ particles is equal to $2.95\,{\%}$). 

Detailed information on the critical strain for the carbon and boron chains for different $N$ is given in Table~\ref{t5} and Table~\ref{t7}.

It is interesting that for carbon chains there is a significant dependence of the critical strain $\eta_{c}$ on the parity of $N/2$, which is described in detail in Sec.~\ref{Seq5}, whereas it is barely present for the boron chains (see Fig.~\ref{fig8}).

\subsection{Comparison of the bi-structure with two long bonds in carbon and boron chains \label{Seq7.3}}

Similar to the case of the bi-structure 1, the critical strain $\eta_{c}(N)$ for the bi-structure 2 decreases rather rapidly with increasing $N$ both for carbon and boron chains (see Table~\ref{t9} and Table~\ref{t13}).

The bi-structure 2 for carbon and boron chains appears at greater values of the critical strain $\eta_{c}$, compared to those for the bi-structure 1. For example, for the carbon chain with $N=25$, $\eta_{c}$ is equal to $5.20\,{\%}$, whereas for close values of $N$ (it must be {\it even} for the existence of the bi-structure 1) $\eta_{c}(N)$ has the following values: $\eta_{c}(24)=3.57\,{\%}$ and $\eta_{c}(26)=2.07\,{\%}$.

Similar critical values $\eta_{c}(N)$ for boron chains with the same $N$ for appearance of the bi-structure 2 are: $\eta_{c}(25)=9.40\,{\%}$ and $\eta_{c}(27)=8.74\,{\%}$, whereas for the bi-structure 1 in the boron chain we have: $\eta_{c}(26)=6.71\,{\%}$.

\subsection{The length of the long bond and interactions between sub-chains of the bi-structures \label{Seq7.4}}

For both types of bi-structures in carbon and boron chains, the long bonds, appearing as a result of the corresponding bifurcation, are significantly larger than the average lengths of the short bonds (usually about 1.5 - 2 times). For example, for bi-structure 1 with the critical strain $\eta_{c}=6.11\,{\%}$ of the carbon chain with $N=16$ atoms, the long bond is $L_{l}=2.58\,\buildrel_\circ \over {\mathrm{A}}$, while directly before the corresponding bifurcation the bond length between the same atoms is $L_{s}=1.38\,\buildrel_\circ \over {\mathrm{A}}$, e.g. the bond length increases 1.87 times. For the chain of $N=16$ boron atoms, the corresponding values are $L_{l}=3.71\,\buildrel_\circ \over {\mathrm{A}}$ and $L_{s}=1.72\,\buildrel_\circ \over {\mathrm{A}}$, which corresponds to increase of the bond length by a factor of 2.16.

For the bi-structure 2 in the chain of $N=17$ carbon atoms the crossing over the critical strain $\eta_{c}=8.10\,{\%}$ leads to the bond length increase from $L_{s}=1.40\,\buildrel_\circ \over {\mathrm{A}}$ to $L_{l}=2.07\,\buildrel_\circ \over {\mathrm{A}}$, and for the same chain of boron atoms from $L_{s}=1.74\,\buildrel_\circ \over {\mathrm{A}}$ to $L_{l}=2.85\,\buildrel_\circ \over {\mathrm{A}}$.

Since long bonds in all bi-structures are significantly longer than short bonds, the overlap of the atomic orbitals decreases drastically as a result of the corresponding bifurcation. Therefore, the interaction between these atoms becomes significantly weaker than that for short bonds.

However, the sub-chains of the given bi-structure are still bound by the attractive van der Waals forces. The presence of attraction between the sub-chains can be proved by the following arguments:
\begin{enumerate} 
    \item When we take two cumulene chains shifted relative to each other along their axis by a distance of $3\,\buildrel_\circ \over {\mathrm{A}}$ as the initial configuration for constructing the carbon bi-structure 1, the distance between these sub-chains {\it decreases} as a result of the DFT refining.
    \item One can observe a complex atomic {\it dynamics} in the vicinity of the bi-structures 1 and 2. It is interesting that in the neighborhood of the bi-structure 1 in the Lennard-Jones model we have found some {\it discrete breathers} (DB) of a new type~\cite{23}. These breathers, in contrast to the well-known DBs by Sievers and Takeno and those by Page, demonstrate monotonous decrease of atomic amplitudes from the center of DB to its periphery. Moreover, it occurs that the sequence of these amplitudes represents an almost exact arithmetic progression. Unfortunately, we failed yet to construct such discrete breathers in the framework of the DFT theory because of some technical problems. This is the goal of further research.
\end{enumerate}

In conclusion, we would like to note that our study of the Lennard-Jones model suggests that some other equilibrium states besides above described bi-structures can exist in the strained carbon and boron chains. This problem will be discussed elsewhere.

\section*{Acknowledgments}

The authors acknowledge support by the Ministry of Science and Higher Education of the Russian Federation (state assignment grant No. 3.5710.2017/8.9) and they are sincerely grateful to N. V. Ter-Oganessian for useful discussions.

\bibliographystyle{unsrt}
\bibliography{New}

\end{document}